# A Stochastic-MILP dispatch optimization model for Concentrated Solar Thermal under uncertainty

Navid Mohammadzadeh[a,*], Huy Truong-Ba[a], Michael E. Cholette[a],
Theodore A. Steinberg[a],Giampaolo Manzolini[b]
[a] School of Mechanical, Medical and Process Engineering, Queensland University of Technology, Australia
[b] Department of Energy, Politecnico di Milano, Italy
[*] Corresponding author, Mechanical, Medical and Process Engineering Department, Queensland University of Technology, 4000 Brisbane, Australia. Email address: n.mohammadzadeh@qut.edu.au

## Abstract

Concentrated Solar Thermal (CST) offers a promising solution for large-scale solar energy utilization as Thermal Energy Storage (TES) enables electricity generation independent of daily solar fluctuations, shifting to high-priced electricity intervals. The development of dispatch planning tools is mandatory to account for uncertainties associated with solar irradiation and electricity price forecasts as well as limited storage capacity. This study proposes the Stochastic Mixed Integer Linear Program (SMILP) to maximize expected profit within a specified scenario space. The SMILP scenario space is generated by different Empirical Cumulative Distribution Function (ECDF) percentiles of the potential solar energy to accumulate in storage and the expected profit is estimated using the Sample Average Approximation (SAA) method. SMILP exhibits robust performance, however, its computational time poses a challenge. Thus, three heuristic solutions are developed which run a set of deterministic optimizations on different historical weather profiles to generate candidate dispatching plans (DPs). The candidate DP with the best *average* performance on all profiles is then selected. The new methods are applied to a case study for a 115 MW CST plant in South Australia. When the historical database has a limited set of historical weather profiles, the SMILP achieves 6% to 9% higher profit than the closest benchmark when the DP is applied to novel weather conditions. With a large historical weather data, the performance of SMILP and closet Heuristic becomes nearly identical due to the fact that the SMILP can only utilize a limited number of trajectories for optimization without becoming computationally infeasible. In this case, the heuristic emerges a practical alternative, since it provides similar average profit in a reasonable time (saving about 7 hours in computing time). Taken together, the results illustrate the importance of considering uncertainty in DP and indicate that straightforward heuristics on a large database are a practical method for addressing uncertainty.

## 1 Introduction

Concentrated Solar Thermal (CST) and Utility-scale Photovoltaic (PV) technologies are the predominant technologies for large-scale solar energy exploitation [1]. CST can easily exploit Thermal Energy Storage (TES) and offer a semi-dispatchable solution with superior regulation characteristics relative to Utility-scale PV that needs batteries with consequent costs penalization. CST allows higher revenue from dispatching during high-price periods [2]–[4], shorter start-up/shutdown times [5], fewer generation unit start-up/shutdowns[6], and minimal issues for grid stability and power system reliability [7]–[9].

Thermal energy generation relies on inherently random solar irradiance which can be overcome by the adoption of large storage, albeit with significant capital expenditure. On the contrary, aggressive storage use may lead to immediate profit, but at the cost of missing future revenue opportunities, especially during high electricity prices and poor solar resource conditions [10]. Overly aggressive use may also accelerate the degradation of critical subsystems due to unplanned shutdowns and unnecessary load changes. This justifies the development of a sophisticated dispatch optimization model that aims to maximize the economic benefits of dispatching when interacting with a stochastic environment.

Dispatch Planning (DP) optimization is an active field of research, particularly given the relatively short history of CST operation. Mathematical Programming, such as Mixed-Integer Linear/Nonlinear Programming (MILP/NMILP), represents a common approach widely used for DP optimization modelling [11]–[13]. Wagner et al. [14] proposed a high-level of detail MILP model for CST plants, aiming to maximize dispatch profit discounted over the optimization horizon through careful utilization of storage and limitation of degradation of critical subsystems via controlling startup/shutdown. This model has since been widely used in studies for dispatching CST plants. Hamilton et al. [15] extended the Wagner's model for the CST plant when it collocated with a Utility-scale PV plant integrated with electrical batteries. The proposed Wagner's optimization model was integrated into the maintenance and failure simulation model in [16] to account the plant downtime caused by normal wear and tear and cycling of the plant.

However, the majority of DP optimization models suggested by the studies above are conducted in a deterministic environment, requiring accurate forecasting for uncertain weather and electricity prices. Yet, the time-varying statistical properties of uncertain parameters make forecasting models notoriously challenging [17]–[21], and achieving accuracy often necessitates ad-hoc parameter tuning by experts [22]. Moreover, the solutions obtained from deterministic optimization tend to be fragile, with forecasting errors leading to considerable deviations from the optimal DP [23][24].

There is a significant body of work that tackles uncertainties in DP optimization through robust [25][26] and interval optimization approaches [27]. The robust optimization is solved for the optimized DP considering the worst-case scenario for uncertain parameters [28]. This assumption yields overly-conservative DPs and the underestimation of profit [25][29]. Interval optimization considers both the worst and best plans within intervals, which results in an inaccurate representation of the expected profit due to the wide optimal ranges [30].

An alternative approach to handling uncertainty is stochastic optimization, where uncertainty is represented by a set of representative scenarios from the full distribution [31]. Some studies have applied stochastic optimization for optimized DP for hybrid CSP-Biomass plants in electricity markets [32], CST generation in spot markets [33], and joint CST-Wind systems in day-ahead and ancillary markets [34]. Despite complexity and computational requirements, stochastic optimization can provide operation schedules that outperform deterministic models [33][35]. However, a drawback is the need for accurate probability distributions to characterize the uncertainty [30]. To estimate probabilistic components affected by uncertainty, Sample Average Approximation (SAA) is widely used [36]–[38]. Lima et al. [39] applied SAA to solve risk-average problems in a virtual power plant scheduling.

An important challenge in applying SAA is the handling of hard constraints (e.g., the limitations on the stored energy). Typically, *chance constraints* are employed, where the hard constraint is replaced with a probabilistic version that permits some violations of some of the samples used in the SAA [40][41]. Further details on Chance-Constrained Program (CCP) can be found in [42][43]. In the CST context, however, there are three important challenges with using CCP. First of all, the appropriate selection of the probability level is challenging; it is often a matter of experience to find a good compromise between constraint satisfaction and profit-seeking [44]. A high probability level may induce an overly conservative DP, while low probability level consequently leads the violation of chance constraints for many scenarios and thus overestimation of expected profit. Yet, the right choice of probability level is not clear and improper settings may lead to an unreliable operation schedule or computational difficulties [45].

In addition, the feasibility of the dispatch solution is not maintained for all trajectories and the SAA could therefore be a poor approximation of the true average objective, leading the results of optimization hard to interpret [45]. Finally, for probability levels close to one, approximating the violation probability requires a large number of scenarios. For instance, the significance level of $\epsilon = 0.01$ (i.e., the probability level of $0.99$) will require at least 100 scenarios to permit any violation. Such a large number of scenarios may drive a prohibitively large computational cost.

The main contribution of this study is to develop a novel Stochastic Mixed Integer Linear Program (SMILP) for a CST plant while avoiding chance constraints. The SAA of the expected profit across all scenarios is maximized. Historical meteorological records are used to establish sample trajectories which serve as scenarios for the SMILP SAA. A case study is conducted for a hypothetical 115 MW CST plant with 8-hour storage in South Australia. The proposed SMILP is benchmarked against an idealized Perfect Knowledge (PK) forecast and three (feasible) heuristic approaches. A simulation model developed in [24] is adopted to evaluate the performance of SMILP against benchmarks under novel conditions. A major advantage of the developed methodology is its versatility such that it is easily adaptable to different CST layouts with TES.

The remainder of the paper is organized as follows. Section 2 details the operating principles of the CST-TES plant, followed by formulation of the proposed SMILP model in Section 33, sampling procedure and benchmarks in Section 4, case study in Section 5, results and discussion in Section 6, and conclusion in Section 7.

## 2 Operation model of CST-TES

Figure 1 shows a simplified block diagram CST-TES with the connections between key subsystems which is considered in this study. Solid lines indicate the input/output power to/from key subsystems. The dashed lines are dispatch control signals. Due to the existence of uncertainty, the plant does not necessarily follow the given DP. The hat ($\hat{\cdot}$) is used to distinguish the control (or "planned") inputs from the "actual" plant's operation.

Two types of control inputs are applied to the system. The first type involves binary variable $\hat{y}_k^{(\cdot)} \in \{0,1\}$ specifying the operating/stopping time for receiver and power block, where $k = 1, .., K$ is the time index, and $K$ is the planning horizon. The second type of control input is continuous variables, providing the system with planned storage charging and discharging rates, denoted by $\hat{q}_k^r \in \mathbb{R}^{\geq 0}$ and $\hat{q}_k^c \in \mathbb{R}^{\geq 0}$, respectively, while the actual storage charting and discharging rates are $q_k^{ract} \in \mathbb{R}^{\geq 0}$ and $q_k^{cact} \in \mathbb{R}^{\geq 0}$, respectively. The CST-TES configuration consists of four key subsystems, i.e., heliostat fields, receiver, molten-salt storage, and power block. The functionality of these subsystems is detailed in the following.

**Figure 1: Concentrated Solar Thermal (CST) with Molten-salt Thermal Energy Storage (TES)**

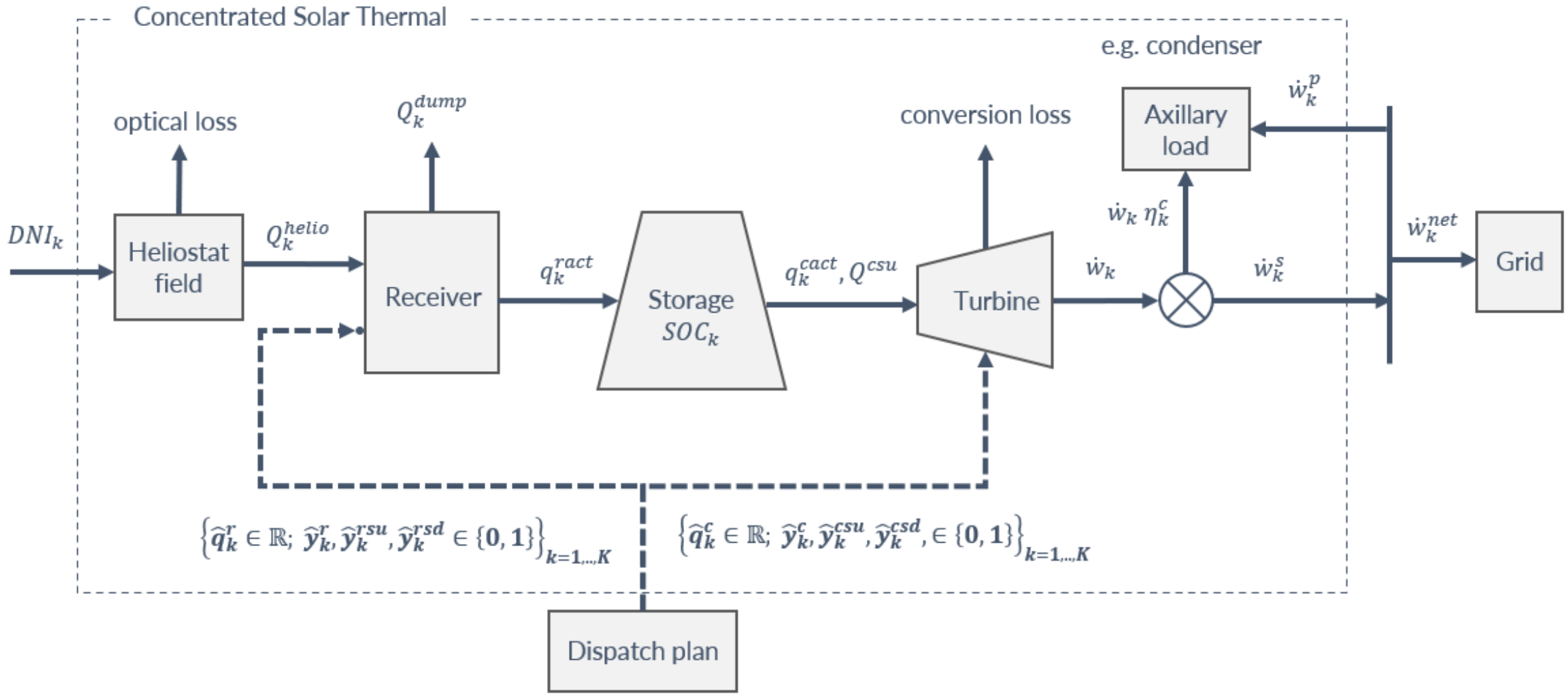

The total thermal power generated via heliostat field is the function of characteristics of heliostat field and solar irradiance, which is denoted by $Q_k^{helio}$ and calculated using Eq. (1):

$$Q_k^{helio} = N^{helio} A^{helio} \rho_k \, c_k \, \eta_k^{opt} \, DNI_k \tag{1}$$

where $N^{helio}$ is the total number of heliostats, $A^{helio}$ is the heliostat's reflective area, $\rho_k$ is the mirror reflectance and soiling coefficient, $c_k \in [0,1]$ is the heliostat availability, $\eta_k^{opt}$ is optical efficiency, and $DNI_k$ is the Direct Normal Irradiation (DNI) at time index $k$.

The reflected solar energy from the heliostats is collected onto the receiver, where a fraction is lost to the ambient through radiative loss i.e., $Q_k^{rad} = f\left(T_k^{amb}\right)$ and convective loss i.e., $Q_k^{conv} = f\left(T_k^{amb}, v_k\right)$, where $T_k^{amb}$ and $v_k$ are ambient temperature and wind speed, respectively. The correlations to calculate these two losses are presented in [14]. A constant loss is assumed for the piping system i.e., $Q^{pipe}$. The potential thermal power for the use in the receiver is denoted by $Q_k^p$ and quantified via Eq. (2):

$$Q_k^p = \max\left[0, Q_k^{helio} - Q_k^{rad} - Q_k^{conv} - Q^{pipe}\right] \tag{2}$$

A part of $Q_k^p$ is used to elevate the receiver temperature during start up and can be partly recovered during the shutdown. The majority of $Q_k^p$ can be potentially delivered to the storage to increment the state of charge. The actual storage charging rate is $q_k^{ract}$, controlled via the planned control input $\hat{q}_k^r$. The heliostat field is sometimes partially defocused and dumps the excessive useful energy to the ambient, see Eq. (3):

$$Q_k^{dump} = \max\left[0, Q_k^p - Q^{ru}\,\delta_k^{rsu} - Q^{rsd}\,\delta_k^{rsd} - q_k^{ract}\right] \tag{3}$$

where $Q_k^{dump}$ is the thermal power dumped to the ambient from the receiver, $Q^{ru}$ and $Q^{rsd}$ is the constant rate of thermal energy consumption during startup and shutdown, respectively. Before the receiver starts charging the storage, a startup should occur and satisfy both minimum startup period and minimum thermal state requirement – this constraint will be discussed in the following sections. Shutdown also incurs an energy cost since the salt must be drained out of the receiver without freezing before thoroughly defocusing the heliostat field. Literature suggests that draining molten salt requires 25% of hourly energy used at minimum receiver generation rate [46]. $\delta_k^{rsu}$ and $\delta_k^{rsd}$ are equal to one if the receiver is in startup and shutdown at time index $k$, respectively, otherwise zero.

The energy accumulated in the storage may be used to run the power block to generate electricity. An intermediate heat exchanger allows heat transfer between storage medium (e.g., molten salt) and steam (or equivalent working fluid) and the steam is used to spin the turbine for electricity generation. Eq. (4) measures the plant's net output $\dot{w}_k^{net}$ that can be potentially dispatched to the grid:

$$\dot{w}_k^{net} = \dot{w}_k\,(1 - \eta_k^c) - \dot{w}_k^p \tag{4}$$

where $\dot{w}_k$ is the rate of electricity generation in the power block, a part of which is consumed locally to respond auxiliaries' consumptions $\eta_k^c$, and $\dot{w}_k^p$ is the rate of electricity purchased from the grid to run the plant's auxiliaries e.g., pumping systems and electrical power to drive heliostat mirrors.

# 3 SMILP dispatch optimization model

The SMILP model for the system shown in Figure 1 is presented in this section. This model aims to optimize the DP such that SAA of expected profit is maximized for a given scenario space – which includes trajectories for uncertain parameters (e.g., DNI)[1]. Several key notations must be defined before outlining the SMILP model.

Once again, let $k$ be the time index, $K$ is the optimization horizon, and each scenario is indexed by $s \in \mathbb{S}$, where $\mathbb{S}$ is the scenario space. The decision variables are grouped into *plan variables* —denoted $(\hat{\cdot})_k$ — and *control variables* — denoted $(\cdot)_{k,s}$. The difference in the subscripts is because the plan variables must be the same for all scenarios whereas the control variables must be scenario-specific to ensure satisfaction of physical limits. This grouping of decision variables enables the definition of linear-logical constraints and the simulation of the interaction of physical constraints with uncertainty in the optimization model.

[1] A procedure for generating scenarios will detail in the following sections.

Decision variables are also categorized into binary or continuous variables. The binary variables are devoted to characterizing the operating mode of key subsystems. The symbol $\hat{y}_k^{(\cdot)}$ denotes plan mode-switching signals and $\delta_{k,s}^{(\cdot)}$ is the actual operating mode. The continuous variables, however, describe input/output power to/from key subsystems and the level of thermal energy (e.g., state of charge). The DP (optimized or not) is in the form of the vector of operation decision for the receiver i.e., $\left\{\hat{q}_k^r \in \mathbb{R};\ \hat{y}_k^r,\ \hat{y}_k^{rsu},\ \hat{y}_k^{rsd} \in \{0,1\}\right\}_{k=1,..,K}$ and for the PB i.e., $\left\{\hat{q}_k^c \in \mathbb{R};\ \hat{y}_k^c,\ \hat{y}_k^{csu},\ \hat{y}_k^{csd} \in \{0,1\}\right\}_{k=1,..,K}$, see Figure 1. The key parameters and variables are listed in Table A.1 and Table A.2, respectively, see Appendix A. The subsequent sub-sections describe the objective function and constraints in the proposed SMILP model.

## 3.1 Objective function

The SMILP model aims to maximize the expected profit from dispatching over all possible values of the uncertain variables (e.g., DNI, electricity prices). This problem, however, is intractable, partly due to the complex time- and cross-correlations of the uncertain quantities. Instead, a representative set of scenarios is constructed by sampling from the historical data and a SAA expected profit is maximized as an estimate of the true expected value. The SAA is denoted is $f^{SAA}$ and calculated using Eq. (5):

$$f^{SAA} = \frac{1}{N_s} \sum_{\forall s \in \mathbb{S}} \sum_{k=1}^{K} \lambda^k \left(f_{k,s}^R - f_{k,s}^{aux} - f_{k,s}^{om}\right) \tag{5}$$

where $N_s$ is the size of scenario space, and $0 < \lambda \leq 1$ is the discount rate. Three terms on the right-hand side are the revenues from selling the electricity $f_{k,s}^R$, the cost of purchasing the electricity from the grid to run the auxiliaries $f_{k,s}^{aux}$, and the total operation and maintenance costs $f_{k,s}^{om}$. This formulation, therefore, has the explicit terms of the revenue and costs. The term $f_{k,s}^R$ is defined as:

$$f_{k,s}^R = \Delta t \cdot p_{k,s} \cdot \dot{w}_{k,s}^s \quad ;\ \forall k\ \wedge \forall s \in \mathbb{S} \tag{6}$$

where $\Delta t$ is the duration of each interval, $p_{k,s}$ is the electricity price, and $\dot{w}_{k,s}^s$ denotes the dispatched power to the grid at time $k$ in scenario $s$. The cost of purchasing electricity from grid, which is denoted as $f_{k,s}^{aux}$ is calculated:

$$f_{k,s}^{aux} = \Delta t \cdot p_{k,s} \cdot \dot{w}_{k,s}^p \quad ;\ \forall k\ \wedge \forall s \in \mathbb{S} \tag{7}$$

where $\dot{w}_{k,s}^p$ is the power purchased from the grid at time $k$ in scenario $s$. The total operation and maintenance cost associated with the key subsystems $f_{k,s}^{om}$ is calculated as:

$$f_{k,s}^{om} = \Delta t \cdot \left(C^{rec} \cdot q_{k,s}^{ract} + C^{c} \cdot \dot{w}_{k,s}\right) + \left(C^{\delta W} \cdot \dot{w}_{k,s}^{\delta} + C^{rsup} \cdot \delta_{k,s}^{rsup} + \alpha \cdot \delta_{k,s}^{rsd} + C^{csup} \cdot \delta_{k,s}^{csup} + \alpha \cdot \delta_{k,s}^{csd}\right) \quad (8)$$

where $C^{rec}$ is the degradation cost for each MWh generation by receiver; $q_{k,s}^{ract}$ is the actual receiver thermal power used to charge the storage; $\dot{w}_{k,s}$ represents the rate of electricity generation by the power block; $C^{c}$ is the degradation cost associated with each unit of electricity generation; $C^{\delta w}$ is the degradation rate for each MW ramping of the power block $\dot{w}_{k,s}^{\delta}$; $C^{rsup}$ and $C^{csup}$ are the degradation costs associated with each startup event in the receiver and the power block, respectively; and $\alpha$ is the monetary cost assigned to the shutdown event. The actual receiver startup and shutdown events are denoted by $\delta_{k,s}^{rsup} \in \{0,1\}$ and $\delta_{k,s}^{rsd} \in \{0,1\}$, respectively. As with receiver, $\delta_{k,s}^{csup} \in \{0,1\}$ and $\delta_{k,s}^{csd} \in \{0,1\}$ represent the actual power block's startup and shutdown, respectively. The objective function is subjected to physical and logical constraints controlling the startup, power generation, and shutdown in each subsystem, which are presented in the proceeding subsections.

## 3.2 Receiver

### 3.2.1 Logic plan constraints

In this model, three operating modes are defined for the receiver, namely start-up, power generation, and off-mode, thus three binary plan variables: 1) $\hat{y}_{k}^{rsup} = 1$ specifying time to drive a startup event; zero otherwise, 2) $\hat{y}_{k}^{r} = 1$ representing the time for the receiver to charge the storage; zero otherwise, and 3) $\hat{y}_{k}^{rsd} = 1$ determining the time planned for the shutdown. In contrast with power generation mode which can happen over multi-time steps, the cold startup and shutdown are events (commands) and occur in a single time step. Constraints (A.1) to (A.5) impose relationships between these three plan variables, see Appendix A.

### 3.2.2 Receiver startup

One of the major challenges in this study is to separate the plan and the actual (secnario-specific) control variables. First, consider the startup process where a part of the thermal power consumes to raise the reciever's thermal state. Constraint (9) drives the thermal state $e_{k,s}^{rsu}$ with a constant rate $Q^{ru}$ during startup. Constraint (10) restricts the thermal state to a certain threshold $E^{r}$, and resets the thermal state post startup completion. The DNI shortage in a (some) scenario(s) may consequently lead to a delay in the receiver's startup. A non-negative, auxiliary variable (i.e., $\phi_{k,s}^{r} \geq 0$) is defined to model the contingencies in the startup procedure in the scenario(s). After startup completion, Constraint (11) allows the activation of the slack variable to respect right and left hand-sides of Constraint (9). Constraint (12) keeps the slack variable equal to zero during startup.

$$e_{k,s}^{rsu} + \phi_{k,s}^{r} = e_{k-1,s}^{rsu} + \Delta t \cdot Q^{ru} \cdot \delta_{k,s}^{rsu} \qquad \forall k \wedge \forall s \in \mathbb{S} \qquad (9)$$

$$e_{k,s}^{rsu} \leq E^{r} \cdot \delta_{k,s}^{rsu} \qquad \forall k \wedge \forall s \in \mathbb{S} \qquad (10)$$

$$\phi_{k,s}^{r} \leq E^{r} \cdot \delta_{k-1,s}^{rsu} \qquad \forall k \wedge \forall s \in \mathbb{S} \qquad (11)$$

$$\phi_{k,s}^{r} \leq E^{r} \cdot (1 - \delta_{k,s}^{rsu}) \qquad \forall k \wedge \forall s \in \mathbb{S} \qquad (12)$$

One of the key features of SMILP model is to produce a DP that is feasible for every given scenario. As such, the plant's operation must obey several multi-statement logical constraints in each scenario. Auxiliary 0-1 switches are defined to describe these conditional statements. Let consider switch $z_{k,s}^{(1)} = 1$ when the potential thermal power $Q_{k,s}^{p}$ falls below the minimum allowable receiver generation limit at a time index $k$ in a scenario $s$; zero otherwise, see Constraint (A.6). Switch $z_{k,s}^{(2)} = 1$ when the thermal state reaches the startup threshold at a time index $k$ in a scenario $s$; zero otherwise, see Constraint (A.7). With $z_{k,s}^{(1)}$ and $z_{k,s}^{(2)}$, a multi-statement function (13) is defined to enforces the receiver's startup at time $k$ in scenario $s$ if all following conditions are met simultaneously:

i) $\hat{y}_{k}^{r} = 1$, i.e., the receiver is planned to operate at time index $k$, and

ii) $z_{k,s}^{(1)} = 0$, i.e., $Q_{k,s}^{p}$ is sufficient for startup at time $k$ in scenario $s$, and

iii) $z_{k-1,s}^{(2)} = 0$, i.e., the receiver's thermal state was below the threshold $E^{r}$ at time $k-1$ and scenario $s$, meaning that the startup has not been completed yet, and

iv) $\delta_{k-1,s}^{r} = 0$, i.e., the receiver was not in power generation in $k-1$ and scenario $s$.

The violation of each of these conditions would lead the startup to delay to a later time. Constraints (A.8) to (A.12) translate these conditions into a set of linear constraints. Constraint (14) ensures that a cold startup event happens at the first step of the startup procedure.

$$\delta_{k,s}^{rsu} = \begin{cases} 1 & \hat{y}_{k}^{r} = 1 \wedge z_{k,s}^{(1)} = z_{k-1,s}^{(2)} = \delta_{k-1,s}^{r} = 0 \\ 0 & \text{otherwise} \end{cases} \qquad \forall k \wedge \forall s \in \mathbb{S} \qquad (13)$$

$$\delta_{k,s}^{rsup} \geq \delta_{k,s}^{rsu} - \delta_{k-1,s}^{rsu} \qquad \forall k \wedge \forall s \in \mathbb{S} \qquad (14)$$

### 3.2.3 Receiver power generation

The potential thermal power $Q_{k,s}^{p}$ can be utilized for different purposes namely to elevate temperature during receiver's startup, to charge the storage, or to drain out the molten-salt from receiver's tubes during shutdown. Constraint (15) ensures the total thermal power consumed for these stages never exceeds $Q^{p}_{k,s}$ at time index $k$ in scenario $s$. A control variable $q^{ract}_{k,s}$ is defined to separate the actual receiver's output from the plan counterpart $\hat{q}^{r}_{k}$ at time index $k$ and scenario $s$. Constraint (16) ensures $q^{ract}_{k,s}$ respects the minimum and maximum rate of generation limits, $Q^{rl}$ and $Q^{rlim}$, respectively. Constraint (17) ensures the plan charging rate respect the minimum and maximum generation rate limits, and $\hat{q}^{r}_{k} = 0$ during cold startup event.

$$q^{ract}_{k,s} + Q^{ru} \cdot \delta^{rsu}_{k,s} + Q^{rsd} \cdot \delta^{rsd}_{k,s} \le Q^{p}_{k,s} \qquad \forall k \ \wedge \forall s \in \mathbb{S} \qquad (15)$$

$$Q^{rl} \cdot \delta^{r}_{k,s} \le q^{ract}_{k,s} \le Q^{rlim} \cdot \delta^{r}_{k,s} \qquad \forall k \ \wedge \forall s \in \mathbb{S} \qquad (16)$$

$$Q^{rl} \cdot \left(\hat{y}^{r}_{k} - \hat{y}^{rsup}_{k}\right) \le \hat{q}^{r}_{k} \le Q^{rlim} \cdot \left(\hat{y}^{r}_{k} - \hat{y}^{rsup}_{k}\right) \qquad \forall k \qquad (17)$$

Constraint (18) quantifies the total available thermal power $Q^{avail}_{k,s}$ after startup and shutdown power requirements deducted from $Q^{p}_{k,s}$. This constraint allows $Q^{avail}_{k,s} = Q^{p}_{k,s}$ unless the receiver is under a startup or a shutdown at time index $k$ in scenario $s$, which subsequently leads to $Q^{avail}_{k,s} < Q^{p}_{k,s}$. An auxiliary 0-1 switch is defined such that $z^{(3)}_{k,s} = 1$ when total available thermal power is inadequate to satisfy the receiver's minimum generation limit, i.e., $Q^{avail}_{k,s} < Q^{rl}$, see Constraint (A.13).

$$Q^{avail}_{k,s} = Q^{p}_{k,s} - Q^{ru} \cdot \delta^{rsu}_{k,s} - Q^{rsd} \cdot \delta^{rsd}_{k,s} \qquad \forall k, \forall s \in \mathbb{S} \qquad (18)$$

With $z^{(3)}_{k,s}$, a multi-statement function (19) is defined to drive the receiver's power generation mode at time $k$ in scenario $s$ if all the following conditions are simultaneously satisfied:

i. $\hat{y}^{r}_{k} = 1$, i.e., the plan is to operate the receiver to charge the storage at time $k$,

ii. $z^{(2)}_{k,s} = 1$ or $\delta^{r}_{k-1,s} = 1$, i.e., the thermal state reaches the threshold at time $k$ in scenario $s$, or the receiver was already under power generating mode at time $k-1$ in scenario $s$,

iii. $z^{(3)}_{k,s} = 0$, i.e., the available thermal power is adequate to run the receiver at least at minimum generation level.

The violation of each of these conditions prevents the receiver from power generation at time $k$ in scenario $s$, thus leads to $\delta^{r}_{k,s} = 0$. Constraints (A.14) to (A.17) translate the conditions in Eq. (19) into linear constraints.

$$\delta_{k,s}^{r} = \begin{cases} 1 & \hat{y}_k^r = 1 \wedge \left(z_{k,s}^{(2)} = 1 \vee \delta_{k-1,s}^{r} = 1\right) \wedge z_{k,s}^{(3)} = 0 \\ 0 & \text{otherwise} \end{cases} \quad \forall k \wedge \forall s \in \mathbb{S} \tag{19}$$

As the thermal power reaching the receiver has a stochastic nature, the actual thermal power at the outlet of receiver, i.e., $q_{k,s}^{ract}$, is scenario specific and random as well. This control input is a function of the plan charging rate $\hat{q}_k^r$, the thermal power reaches the receiver and available to deliver to the storage $Q_{k,s}^{avail,gen}$. With Constraint (20), $Q_{k,s}^{avail,gen} \leq 0$ when $\delta_{k,s}^{r} = 0$, meaning that there is no power available for the receiver to charge the storage.

$$Q_{k,s}^{avail,gen} = Q_{k,s}^{p} \cdot \delta_{k,s}^{r} - Q^{ru} \cdot \delta_{k,s}^{rsu} - Q^{rsd} \cdot \delta_{k,s}^{rsd} \quad \forall k \wedge \forall s \in \mathbb{S} \tag{20}$$

A multi-statement function (21) expresses the relationship between $q_{k,s}^{ract}$, $\hat{q}_k^r$ and $Q_{k,s}^{avail,gen}$. Accordingly, $q_{k,s}^{ract}$ must pursue dispatch setpoint $\hat{q}_k^r$ at time $k$ in scenario $s$ if $Q_{k,s}^{avail,gen}$ is sufficient; the receiver delivers as much power as it can i.e., $Q_{k,s}^{avail,gen}$ to the storage, otherwise. The methodology proposed by [47] is employed to linearize the multi-statement function, see Constraints (A.18) to (A.21).

$$q_{k,s}^{ract} = \begin{cases} 0 & Q_{k,s}^{avail,gen} < Q^{rl} \\ Q_{k,s}^{avail,gen} & Q^{rl} \leq Q_{k,s}^{avail,gen} < \hat{q}_k^r \\ \hat{q}_k^r & Q_{k,s}^{avail,gen} \geq \hat{q}_k^r \end{cases} \quad \forall k \wedge \forall s \in \mathbb{S} \tag{21}$$

### 3.2.4 Receiver shutdown

It is assumed that shutdown event occurs in the last step in the power generating mode, such that it moves the receiver to off mode in the next step, see Constraints (A.22) to (A.25).

## 3.3 Power Block

### 3.3.1 Logical plan constraints

As with the receiver, three modes are considered in the power block's operation, namely startup, power generation, and off-mode, which are controlled by: 1) $\hat{y}_k^{csup} = 1$, specifying the plan time to drive a cold startup at time $k$; zero otherwise, 2) $\hat{y}_k^{c} = 1$, representing the plan time to generate electricity; zero otherwise, and 3) $\hat{y}_k^{csd} = 1$, denoting a plan shutdown at time $k$; zero otherwise. As with the receiver, the cold startup and shutdown are single-time events, while the power generation can take over multiple-time steps. Constraints (A.26) to (A.29) impose the relationship between these plan variables, see Appendix A.

### 3.3.2 Power block startup

The startup procedure in the power block depends on the storage state of charge. However, the state of charge is random as it is a function of the receiver's generation. Constraint (22) allows a linear increase in the thermal state during startup with constant power $Q^c$. The random storage may cause a delay in the startup. A non-negative slack variable $\phi^c_{k,s} \geq 0$ is defined to simulate the potential delay in the startup which results from insufficient storage. Constraints (23) and (24) ensure the slack variable activates post-startup. Constraint (25) limits the thermal state to a specific threshold $E^c$, and resets to zero after startup completion. A 0-1 switch $z^{(6)}_{k,s}$ is defined such that $z^{(6)}_{k,s} = 1$ when the thermal state reaches the threshold at time $k$ in scenario $s$, zero otherwise, see Constraint (A.30). The storage must be sufficient to drive the startup for the power block. Constraint (26) measures the available leave of storage $\varphi_{k,s}$ at time $k$ in scenario $s$. A 0-1 switch $z^{(7)}_{k,s}$ is also defined such that it activates when storage is insufficient to meet startup requirement at time $k$ in scenario $s$, see Constraint (A.31) in Appendix.

$$e^{csu}_{k,s} + \phi^c_{k,s} = e^{csu}_{k-1,s} + \Delta t \cdot Q^c \cdot \delta^{csu}_{k,s} \qquad \forall k \ \wedge \forall s \in \mathbb{S} \qquad (22)$$

$$\phi^c_{k,s} \leq E^c \cdot \delta^{csu}_{k-1,s} \qquad \forall k \ \wedge \forall s \in \mathbb{S} \qquad (23)$$

$$\phi^c_{k,s} \leq E^c \cdot (1 - \delta^{csu}_{k,s}) \qquad \forall k \ \wedge \forall s \in \mathbb{S} \qquad (24)$$

$$e^{csu}_{k,s} \leq E^c \cdot \delta^{csu}_{k,s} \qquad \forall k \ \wedge \forall s \in \mathbb{S} \qquad (25)$$

$$\varphi_{k,s} = s_{k-1,s} - s_{min} + \Delta t \cdot q^{ract}_{k,s} \qquad \forall k \ \wedge \forall s \in \mathbb{S} \qquad (26)$$

Constraint (27) drives the power block's startup provided that all the following conditions are simultaneously satisfied:

i. $\hat{y}^c_k = 1$, i.e., the plan is to operate the power block at time index $k$, and
ii. $\delta^c_{k-1,s} = 0$, i.e., the power block was not in power generating mode at time index $k-1$ in scenario $s$, and
iii. $z^{(6)}_{k-1,s} = 0$, i.e., the startup was incomplete at time index $k-1$ in scenario $s$, and
iv. $z^{(7)}_{k,s} = 0$, i.e., the storage is sufficient for power block's startup at time $k$ in scenario $s$.

The violation of one of these conditions leads to the startup delay to later time. Constraints (A.32) to (A.36) are the linearized representation of above conditions, see Appendix A.

Constraint (28) ensures the cold startup event occurs at the beginning of the power generating mode.

$$\delta_{k,s}^{csu} = \begin{cases} 1 & \hat{y}_k^c = 1 \wedge (\delta_{k-1,s}^c = z_{k-1,s}^{(6)} = z_{k,s}^{(7)} = 0) \\ 0 & \text{otherwise} \end{cases} \quad \forall k \wedge \forall s \in \mathbb{S} \tag{27}$$

$$\delta_{k,s}^{csup} \geq \delta_{k,s}^{csu} - \delta_{k-1,s}^{csu} \quad \forall k \wedge \forall s \in \mathbb{S} \tag{28}$$

### 3.3.3 Power block electricity generation

There are two key dispatch signals controlling electricity generation at time $k$: 1) $\hat{y}_k^c$ which specifies the plan time for electricity generation, and 2) $\hat{q}_k^c$ which describes the plan rate of storage discharge for electricity generation. Constraint (29) allows the power generation (i.e., $\delta_{k,s}^c = 1$) after startup completion (i.e., $z_{k,s}^{(6)}$) or in the continution of power generation from the previous step (i.e., $\delta_{k-1,s}^c = 1$). Constraint (30) limits the plan rate of storage discharge to power block's minimum and maximum generation constraints. This constraint enables $\hat{q}_k^c$ to be equal to startup power $\mathrm{Q}^c$ at cold startup time. Constraint (31) limits the actual rate of discharge $q_{k,s}^{cact}$ to the power block's minimum and maximum limit at time index $k$ in scenario $s$. Constraint (32) allows the total thermal power consumed in the power block for both startup and generation are equal to the setpoint $\hat{q}_k^c$, unless the storage is insufficient which leads to a forced shutdown at time $k$ in scenario $s$.

$$\delta_{k,s}^c \leq z_{k,s}^{(6)} + \delta_{k-1,s}^c \quad \forall k \wedge \forall s \in \mathbb{S} \tag{29}$$

$$Q^l \cdot \left(\hat{y}_k^c - \hat{y}_k^{csup}\right) + \mathrm{Q}^c \cdot \hat{y}_k^{csup} \leq \hat{q}_k^c \leq Q^u \cdot \left(\hat{y}_k^c - \hat{y}_k^{csup}\right) + \mathrm{Q}^c \cdot \hat{y}_k^{csup} \quad \forall k \wedge \forall s \in \mathbb{S} \tag{30}$$

$$Q^l \cdot \delta_{k,s}^c \leq q_{k,s}^{cact} \leq Q^u \cdot \delta_{k,s}^c \quad \forall k \wedge \forall s \in \mathbb{S} \tag{31}$$

$$\hat{q}_k^c - \mathbb{M} \cdot \left(1 - \delta_{k,s}^c\right) \leq q_{k,s}^{cact} + Q^c \cdot \delta_{k,s}^{csu} \leq \hat{q}_k^c \quad \forall k \wedge \forall s \in \mathbb{S} \tag{32}$$

Constraints (A.37) to (A.41) represent the linear correlation to measure the electricity generation in the power block and the electricity purchased from the grid to run auxiliaries.

### 3.3.4 Power block shutdown

The power block shutdown is because of a plan shutdown from DP, or an unplan shutdown due to insufficient storage. Constraints (A.42) to (A.44) control the power block shutdown, see Appendix.

### 3.4 Energy balance in storage

The storage dynamics are scenario-dependent because of the random charging rate. Constraint (33) measures the level of storage at time $k$ in scenario $s$, which is denoted as $s_{k,s}$:

$$s_{k,s} = s_{k-1,s} + \Delta t \cdot \left(q_{k,s}^{ract} - q_{k,s}^{cact} - Q^c \cdot \delta_{k,s}^{csu}\right) \qquad \forall k \ \wedge \forall s \in \mathbb{S} \qquad (33)$$

$$SOC_{min} \cdot E^u \leq s_{k,s} \leq E^u \qquad \forall k \ \wedge \forall s \in \mathbb{S} \qquad (34)$$

where $\Delta t$ is the duration of each interval. Constraint (34) applies lower and upper bounds to storage level, where $SOC_{min}$ is the minimum allowable state of charge and $E^u$ is the storage capacity. Notably, this study assumes a constant power density for storage regardless of the level of energy in storage. In practice, the power density for heat addition is a function of storage level. As such, power density becomes smaller particularly when the storage level positions in the proximity of the upper bound. However, this level of detail is out of the scope of this study.

## 4 Stratified sampling and benchmarks

The DP optimization includes the expectation over several key exogenous variables, e.g., solar irradiation, ambient temperature, wind speed, and electricity prices. In principle, the SAA in Eq. (5) requires a large number of samples of these exogenous variables to be an accurate approximation of the expected profit. However, a large number of samples will lead to an increase in the scenario-dependent variables (e.g., $z_{k,s}^{(\cdot)}$) and thus increase the computational time. It is therefore important to intelligently generate scenarios from the historical database to accurately estimate the expected profit while minimizing the computational burden.

In addition, the seasonality as well as both auto- and cross-correlated weather characteristics (i.e., solar irradiation, ambient temperature, and wind speed) must be preserved in the scenarios. To this end, each day within a given month is assumed to have statistically identical weather properties. The scenarios are generated by sampling the entire time series of weather variables for a selected day (from midnight to midnight) in the historical database. This method allows the preservation of the intra-day correlation while neglecting inter-day correlations.

Consider the case where $N_{year}$ years of data are available in the historical record, which has $N_{scenario}$ two-day sequences in the month of interest[2]. Let $\mathbb{W} = \{1,2,\dots,N_{scenario} \times N_{year}\}$ denote the collection of indices available from the historical database. This set is partitioned into two different sets: a sampling set $\mathbb{W}^{sampling}$ and testing set $\mathbb{W}^{testing}$, such that $\mathbb{W} = \mathbb{W}^{sampling} \cup \mathbb{W}^{testing}$ and $\mathbb{W}^{sampling} \cap \mathbb{W}^{testing} = \emptyset$. The set $\mathbb{W}^{sampling}$ is used for the scenario generation procedure, whereas $\mathbb{W}^{testing}$ is used to assess the performance of the DPs under novel weather conditions. Similarly, electricity prices would also need to be sampled, however, this work assumes that the plant dispatches under a contractual agreement (e.g., a two-tier price scheme). Therefore, the only uncertainty considered in this study is associated with weather variables.

Different scenario-reduction techniques have been suggested in the literature. For instance, Ref. [48] adopted the k-means algorithm [49] to cluster weather historical records and used the moment matching method [50] to select representative scenarios. However, this technique seeks the central of each cluster based on averaging. This method would lead the representative scenario to not necessarily exist in the historical records. In this paper, an Empirical Cumulative Distribution Function (ECDF) is built on the maximum solar energy that could be collected (less losses) for each $s \in \mathbb{W}^{sampling}$, which is denoted as $E_s^{in}$:

$$E_s^{helio} = \Delta t \cdot \rho_t \cdot c \cdot N^{helio} \cdot A^{helio} \cdot \sum_{t=1}^{T} \bar{\eta}_t^{opt} \cdot DNI_{s,t} \tag{35}$$

$$E_s^{in} = max\left[0, E_s^{helio} - E_s^{rad} - E_s^{conv}\right] \tag{36}$$

where $s$ is the index of scenario in $\mathbb{W}^{sampling}$, $t$ is the index of time of day, $E_s^{helio}$ is the total potential solar energy generated via heliostat field in scenario $s$; $\rho_t$ is the mirror reflectance and soiling coefficient in the interval; $c \in [0,1]$ is the heliostat availability; $N^{helio}$ is the total number of heliostat mirrors; $A^{helio}$ is the reflective area per heliostat; $\bar{\eta}_t^{opt}$ is the average optical efficiency which is a function of the time of year and day; $DNI_{s,t}$ is the direct normal irradiation in scenario $s$ and time $t$. $E_s^{in}$ is less than $E_s^{helio}$ due to radiative loss — i.e., $E_s^{rad} = f\left(T_s^{amb}\right)$ — and convective loss — i.e., $Q_s^{conv} = f\left(T_s^{amb}, v_s\right)$, where $T_s^{amb}$ and $v_s$ is the time series associated with ambient temperature and wind speed in scenario $s$, respectively. The correlations to calculate potential radiative and convective loss can be found in [14]. Figure 2 shows an example ECDF for January.

[2] Samples of two-day sequences are drawn to preserve the inter-day correlation of weather variables for the entire optimization horizon. If the optimization horizon is changed, the length of these samples would be adjusted in an obvious way.

**Figure 2: ECDF of the total potential solar energy to store, January, South Australia**

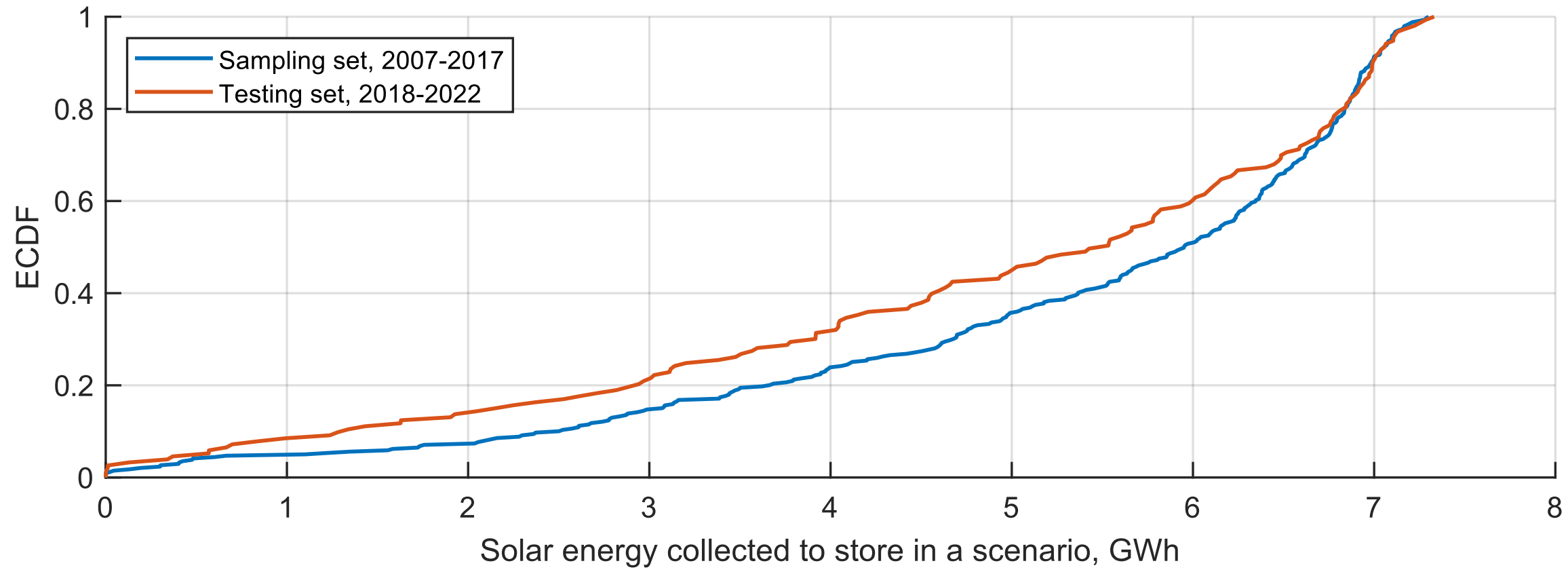

A stratified sampling is used to evenly split the ECDF into $N_s$ percentiles. As each scenario contains two days, a "sample" contains a 48-hour trajectory randomly drawn from each percentile. The output of this procedure is a sampling space $\mathbb{W}_{N_s}^{SMILP} \subseteq \mathbb{W}^{sampling}$ with the size of $N_s$ that is used for the SAA in the SMILP in (5).

## 4.1 Deterministic benchmarks

The solution of SMILP is evaluated against approaches used as benchmarks. The first, named Perfect Knowledge (PK), assumes that all future weather variables are known perfectly. The expected profit obtained from PK is the upper bound since it is not realistic to obtain perfect predictions of weather variables. Subsequently, three heuristic benchmarks are defined:

- *Heuristic-1* uses deterministic optimization to produce candidate DP solutions [51]. A deterministic optimization for each scenario within $\mathbb{W}_{N_s}^{SMILP}$, which generates candidate $DP_s$, $s = 1,2,\dots,N_s$. Subsequently, each $DP_s$ is simulated on $s \in \mathbb{W}_{N_s}^{SMILP}$ to compute its profit on other scenarios $P_{s,s'}$ $\forall s' \in \mathbb{W}^{SMILP}$. These profits are then averaged:

$$SAA_s = \frac{1}{N_s} \sum_{s' \in \mathbb{W}^{SMILP}} P_{s,s'} \tag{37}$$

to obtain and estimate of the expected profit for each of the candidate DPs. The selected DP index is selected as $\hat{s} = \arg\max_s\{SAA_s\}$, see Figure 3.

- *Heuristic-2* is similar to heuristic-1 but uses the whole sampling set $\mathbb{W}^{sampling}$ in place of $\mathbb{W}_{N_s}^{SMILP}$.

- *Heuristic-3* extracts a "typical" day from $\mathbb{W}^{sampling}$ for a particular month via a partitional clustering algorithm (k-medoids [52]) for the month of interest. A DP is optimized using the k-medoid profile for each of the two future days, see [24].

**Figure 3: Heuristic-1 and Heuristic-2 benchmarks**

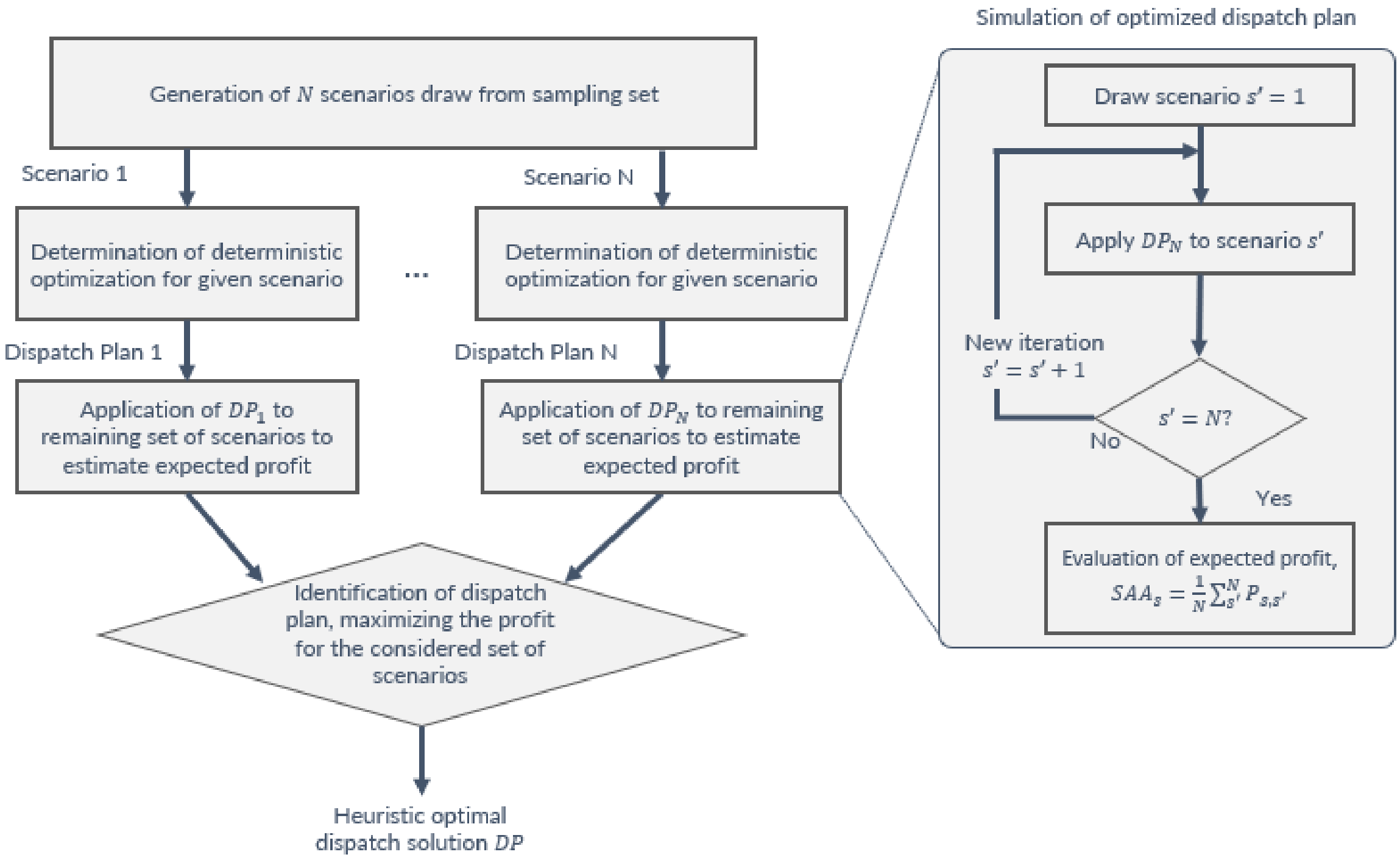

## 4.2 Assessment of DPs under unforeseen scenarios

The performance of heuristics is examined on different weather trajectories from the historical database. In particular, the performance of these benchmarks on $\mathbb{W}^{testing}$ will examine the profit under the realistic scenario that trajectories did not exist in the historical database at the time of optimization. In principle, all heuristics produce DP (e.g., $\hat{q}_k^r$ ) that need to respect the physical constraints – e.g., Constraint (17). This can be considered as computing the auxiliary variables (e.g., $q_{k,s}^{ract}$) in SMILP constraints quantifying the difference between plan and actual variables – e.g., Constraint (21). For the purpose of this work, however, a simulation of the plant is run as the simulation described [24], incorporated all the physical constraints that are present in SMILP.

# 5 Case study: Plant characteristics, costs, and historical data

A case study is conducted, and optimized DPs applied to a hypothetically CST-TES plant in South Australia. System Advisor System (SAM) [53] is employed to extract the key plant's design characteristics, consisting of a 700 MW receiver, a 115 MW steam turbine with 35% thermal-to-electrical conversion efficiency, the solar multiple is 2.13, and 10 hours of thermal storage. The solar field configuration and size of heliostats are optimized in SAM, and plant specifications are reported in Table 1.

**Table 1: Design characteristics of hypothetical ST plant**

| Subsystem | Parameter | Unit | Value |
|---|---|---|---|
| Power block | Design gross power | Mwe | 115 |
| | Maximum thermal input | MWt | 329 |
| | Minimum thermal input | MWt | 65.7 |
| | Start-up energy consumption | MWht | 164.3 |
| Receiver | Maximum generation limit | MWt | 700 |
| | Minimum generation limit | MWt | 175 |
| | Start-up energy consumption | MWht | 175 |
| | Minimum start-up time | Hr | 1 |
| | Hight × diameter | m2 | 19x18 |
| | Tower height | M | 205 |
| Heliostat | Width x length | m2 | 11.3x10.4 |
| | Reflectivity | - | 0.95 |
| | Numbers | - | 11547 |
| Storage | Capacity | Hr | 10 |

Due to the relatively short operating history of CST plants in Australia as well as the sensitive nature of operating data, failure data and detailed system information have limited availability. This is a barrier to a better understanding of the true degradation costs for key subsystems in CST-TES plants [54]. However, the operation and degradation costs are key inputs to the optimization program to quantify the objective function. This study adopts the operating and degradation costs reported in existing dispatch optimization studies [15], by proportionally scaling them by the size of each subsystem. The cycling cost is from [55], see Table 2.

**Table 2: Operation and degradation costs**

| Subsystem | Costs | Unit | Value |
|---|---|---|---|
| Power Block | Power generation | $\$/MWht$ | 1.7 |
| | Cold start-up | $\$/start$ | 5451 |
| | Ramp up/down | $\$/\Delta MWe$ | 0.59 |
| Receiver | Power generation | $\$/MWht$ | 3.7 |
| | Cold start-up | $\$/start$ | 7000 |

In general, TES offers CST the opportunity to participate in different electricity markets – i.e., day-ahead [56], wholesale spot market [57], and ancillary services market [58]. However, dispatching in these markets is associated with a high risk of missing revenue. This leads CST plant operators sometimes hedging the revenue generation against the variability and fluctuation of intraday via dispatching under contractual agreements. This study examines the DP optimization assuming a two-tier contracted price profile for electricity sales.

Historical records with half-hourly data for key weather variables (i.e., DNI, ambient temperature and wind speed) for Woomera – a regional area in South Australia where the plant is assumably located – are obtained from an online database SolCast [13]. From these three timeseries, DNI is exploited to quantify the potential solar field generation. The historical data (2007-2017) is used as a sampling dataset for scenario generation, while the data (2018-2022) is used as a testing set as novel weather data for simulation.

It should be noted that there are key assumptions considered in DP optimization. First, it is assumed the storage is set at a lower bound at the beginning of the optimization to cancel the initial economic value of non-empty storage that could allow electricity generation without a whole system start-up. Secondly, it is assumed the receiver and the power block are set in off-mode at the beginning of the optimization. This means a cold startup is required to restart the subsystem's operation. Finally, as deep discharging of thermal storage has an adverse impact on its useful lifetime, a threshold is considered for the lower bound of the state of charge. This manuscript assumes the lower bound for thermal storage is 10% [24].

The SMILP model is written in MATLAB version 2022b [59] and solved using Gurobi solver version 9.5 [60]. Hardware architecture to generate solutions consists of the 64-bit operating system with a processor intel Core i7-8700 at 3.20GHz, running on Windows 11 with 16GB of installed RAM, and 1TB SSD hard drive.

# 6 Results and discussion

In this section, the results are presented and discussed for an illustrative example where the proposed SMILP with a scenario space including ten scenarios (i.e., $\mathbb{W}_{N_s=10}^{SMILP}$) is solved and performance tested against deterministic optimization. Then, the performance of SMILP – with a larger scenario space $\mathbb{W}_{N_s=14}^{SMILP}$ – and all benchmarks are evaluated when the plant is impacted by weather profiles within one of the three categories, i.e., scenario space $\mathbb{W}_{N_s=14}^{SMILP}$, sampling set $\mathbb{W}_{N=339}^{sampling}$, and testing set $\mathbb{W}_{N=153}^{testing}$. Finally, a sensitivity analysis is performed on the size of scenario space provided to the most competitive heuristic, ensuring the plant using the heuristic achieves the same profit as that obtained via SMILP.

## 6.1 Illustrative example

The proposed SMILP is solved to maximize the SAA of expected profit for a scenario space $\mathbb{W}_{N_s=10}^{SMILP}$. The subplots in Figure 4 presents the potential thermal power to store in each scenario, computed using Eq. (2). Two-tier pricing structure is the highlighted area in Figure 5. It is assumed that the receiver and the power block are in off mode and the storage is set to its lower bound at the beginning of the optimization.

### 6.1.1 Dispatch planning via SMILP

This section presents the results and discussion of an illustrative example for the proposed SMILP model. Figure 4 and Figure 5 shows the plan and actual control inputs for the receiver and the power block, respectively. As is evident, the DP is consistent for all scenarios (see blue dashed line, whereas the actual control inputs are scenario-specific and calculated within the optimization (see black solid lines). Horizontal dashed lines show minimum and maximum generation limits of the corresponding subsystem. The state of charge in each scenario is shown via dotted line.

As seen in Figure 4, the receiver is planned to start up at 9:00 on both days and to shut down at 17:00 and 18:30 on the first and second day, respectively. Inadequate solar resources may consequence a delay in actual startup, e.g., scenario 1 day 2. Following the startup completion, the receiver would potentially be able to charge the storage. The actual charging rate $q_{k,s}^{ract}$ must follow the planned rate $\hat{q}_k^r$. This condition holds if the potential thermal power is sufficient to meet the expected charging rate. The charging will continue albeit at a lower than plan charging rate, otherwise. If the thermal power reaching the receiver (after thermal losses from the receiver deducted) drops below the minimum receiver generation limit, the receiver's operation stops with a forced shutdown, see scenario 4 day 2. Shutdown may be also due to plan shutdown provided by DP, see scenario 7. At each shutdown event, thermal energy is consumed to cool down the tubes and transition to off mode.

Figure 5 shows the results corresponding to the power block. The left y-axis is electricity price, and the right y-axis shows the storage SOC and load on the power block. As seen, the DP specifies a start up at 9:30 on the first day, power generation at full capacity during high price interval, and reduce the load to minimum operating load (i.e., 20%) overnight. The actual thermal power provided to the power block in different time-of-day is a function of the plan discharge rate $\hat{q}_k^c$ and the state of charge. The storage discharges at $Q^c$ during startup unless insufficient storage leads to a delay in startup – see scenario 1 where the starts is postponed for one hour. After startup completion, the storage must discharge at $q_k^c$ for electricity generation, unless storage is insufficient, or discharging does not respect the storage lower bound. In the case of one of these two happens, a forced shutdown is triggered (see scenario 1 day 1 at 14:00) which results in halting electricity generation until the receiver replenishes the storage, see scenario 1 day 2 at 13:00. The results demonstrate that the SMILP solution is feasible for every scenario and ensures that storage adheres to its physical constraints.

**Figure 4: The receiver's plan and control variables under SMILP**

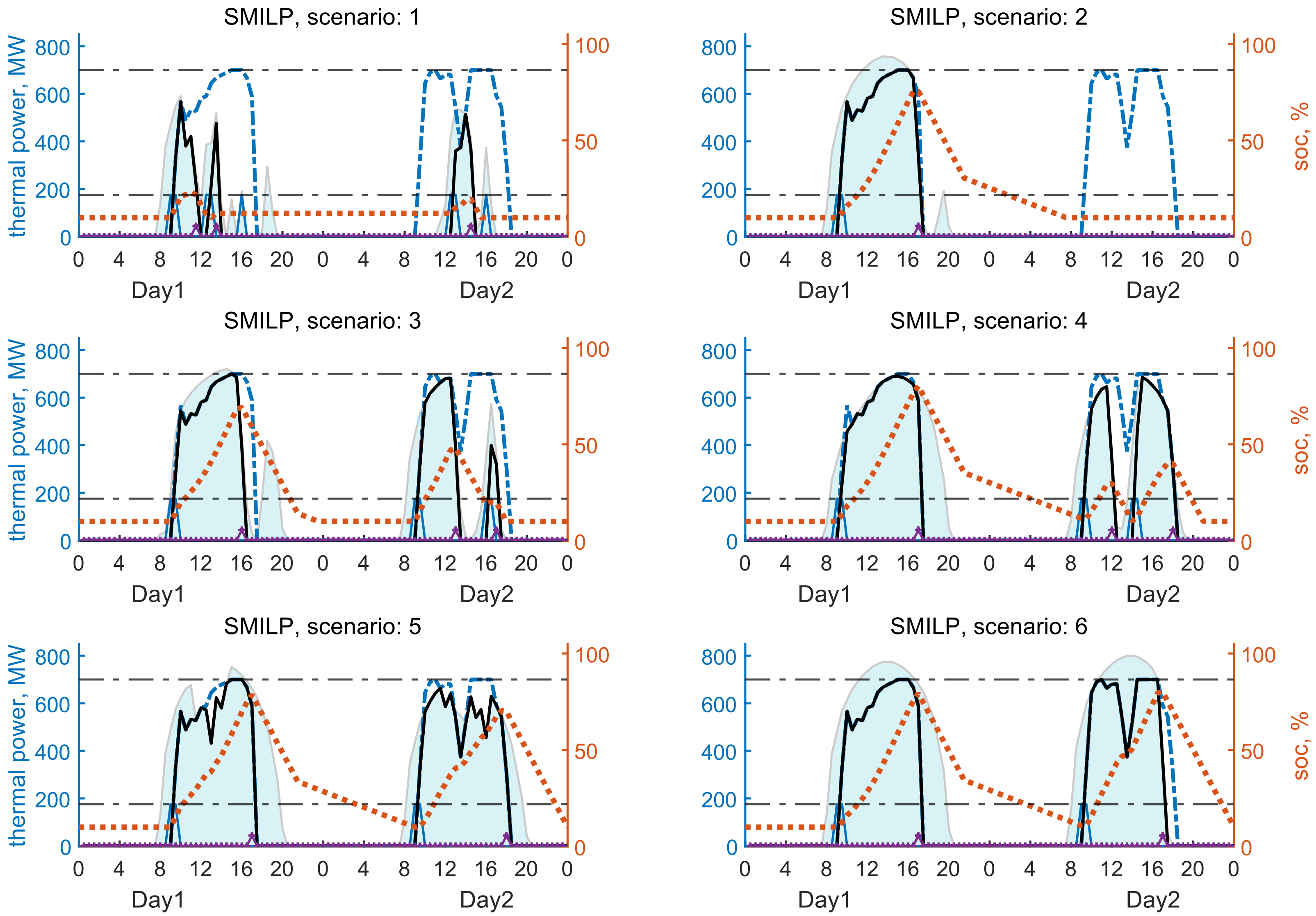

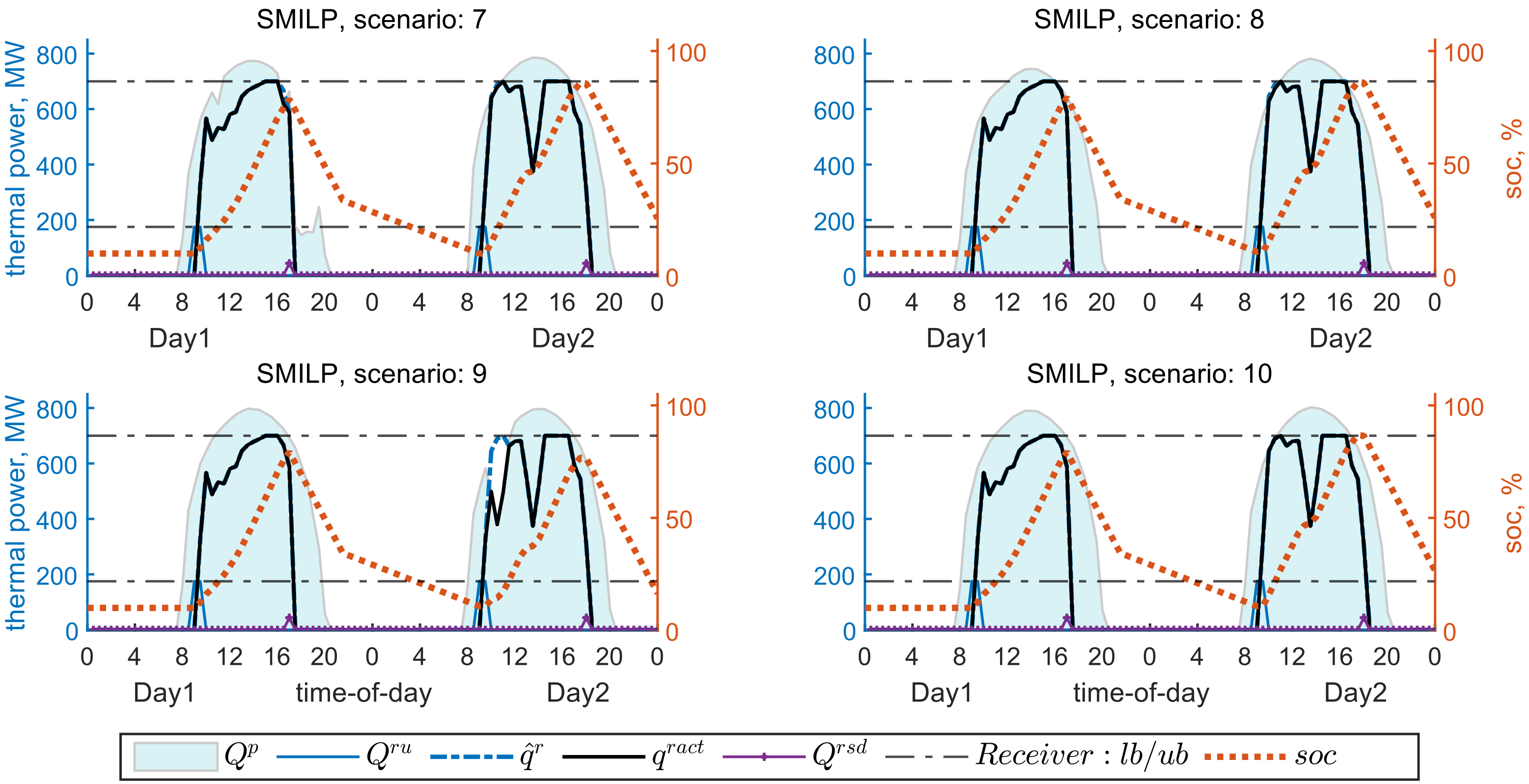
SMILP, scenario: 7
SMILP, scenario: 8
SMILP, scenario: 9
SMILP, scenario: 10
thermal power, MW
soc, %
800
600
400
200
0
100
50
0
0 4 8 12 16 20 0 4 8 12 16 20 0
Day1
time-of-day
Day2
$Q^p$
$Q^{ru}$
$\hat{q}^r$
$q^{ract}$
$Q^{rsd}$
$Receiver : lb/ub$
$soc$

**Figure 5: Plan and control variables of Power Block (PB) for dispatching using SMILP**

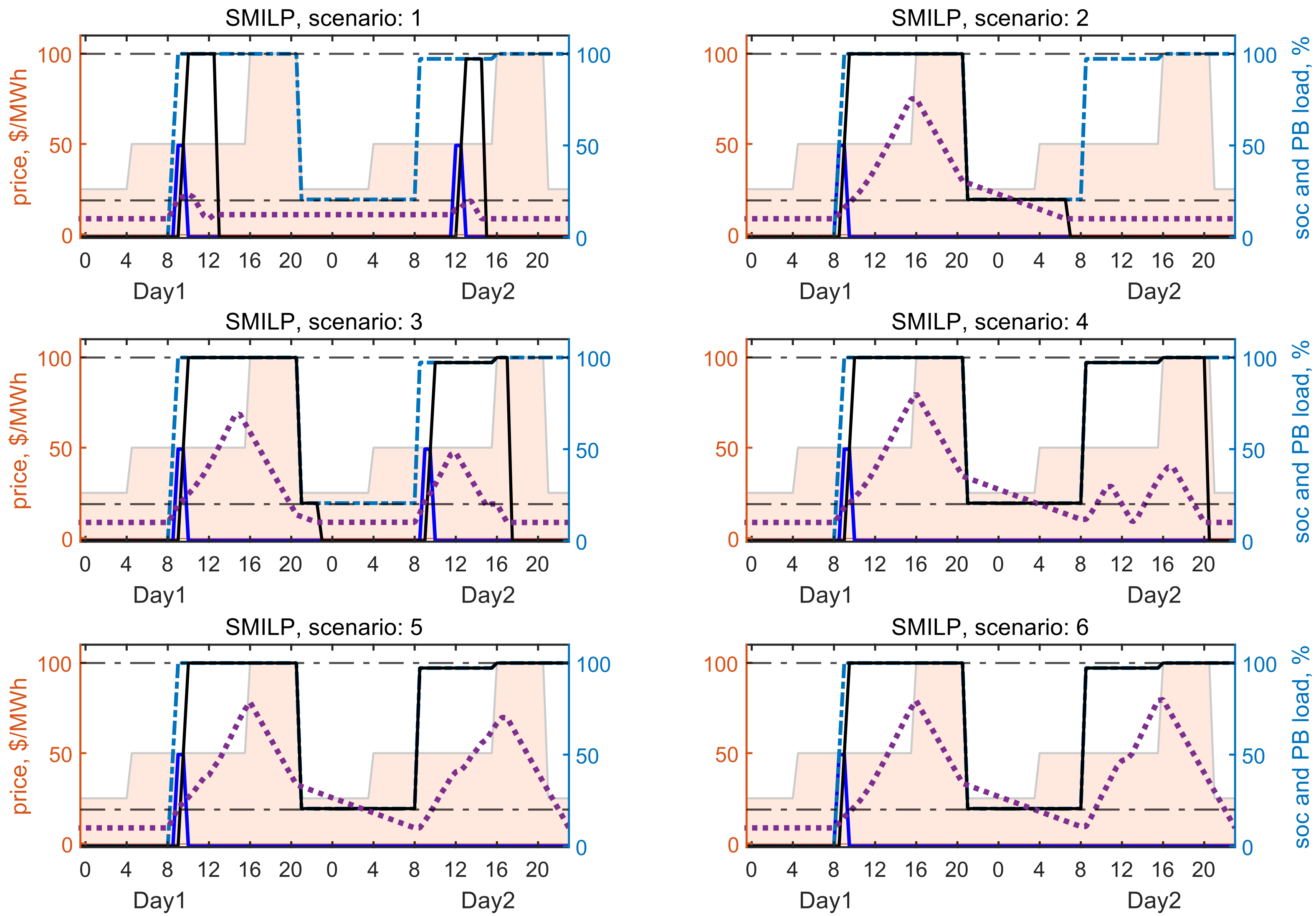

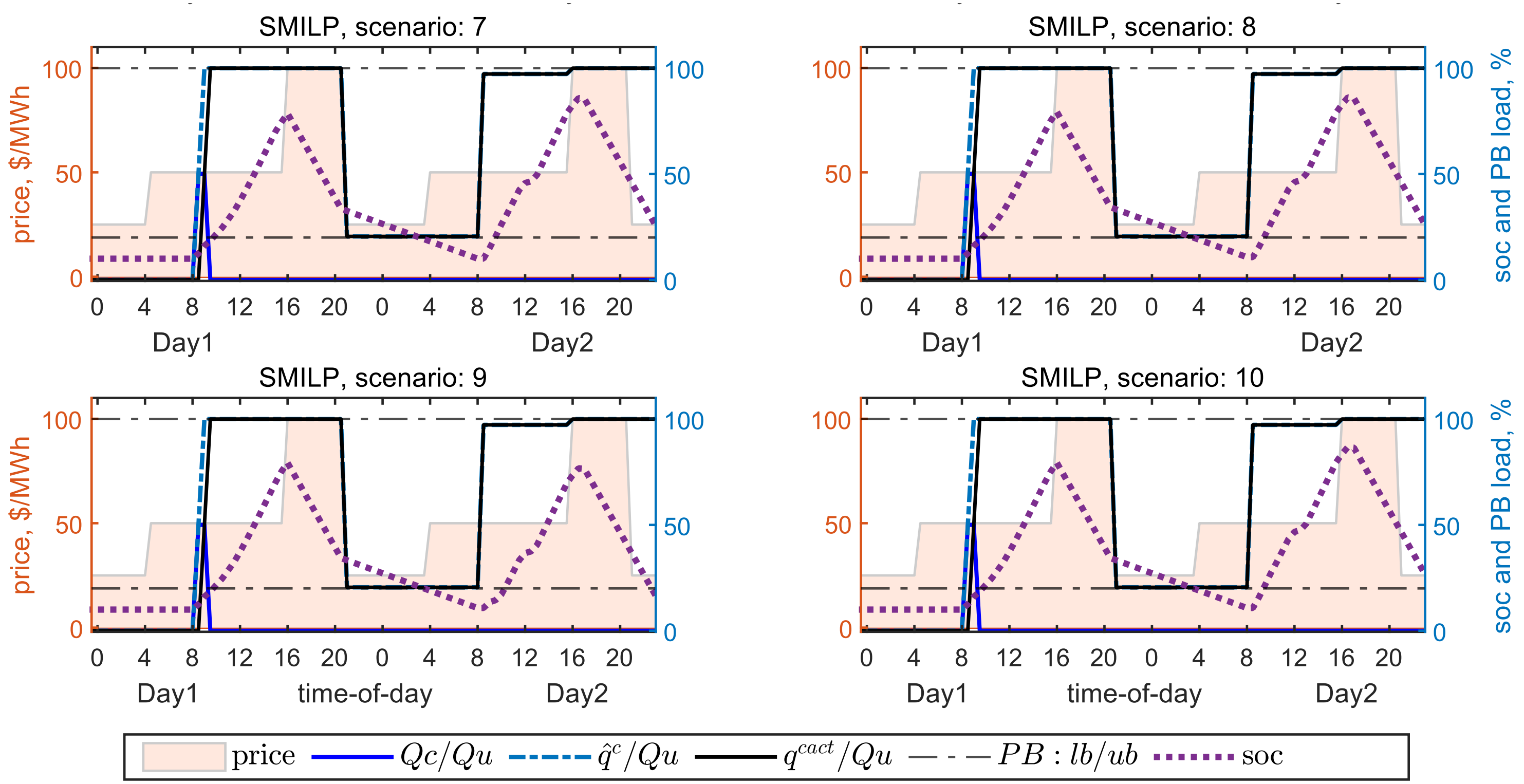

SMILP, scenario: 7
SMILP, scenario: 8
SMILP, scenario: 9
SMILP, scenario: 10
price, $/MWh
soc and PB load, %
0
50
100
0 4 8 12 16 20 0 4 8 12 16 20
Day1
time-of-day
Day2
price
Qc/Qu
q̂^c/Qu
q^cact/Qu
PB : lb/ub
soc

0

### 6.1.2 Dispatch planning using Heuristic-1

This section presents the results of Heuristic-1. Figure 6 shows the optimized DP and calculated the actual variables in the receiver's operation. In this case, the DP tends to extend the receiver operation until 19:00 on both days – which is 2.5 and 1.5 hours longer in day one and two compared to SMILP, respectively. This is likely because the DP in Heuristic-1 is derived from a scenario with two clear sky profile, e.g., scenario 10, and obtains highest SAA when applied to other scenarios. This results in an increase the forced shutdown event at the end of the day in some scenarios. In this dispatching schedule, the average number of receiver cold startups increases by 8% and forced shutdowns by 4.5% compared to SMILP.

A closer examination exhibits that Heuristic-1 intends to operate the receiver at full capacity on the first day but restricts it to 47% capacity from 9:00 to 13:00 on the second day. This increases the risk of empty storage if solar irradiation is insufficient to charge the storage after 13:00, resulting in missing revenue from dispatching during evening peak prices, as observed in scenarios 3 and 4.

Figure 7 shows the results for the power block. As Heuristic-1 is overly optimistic about solar resources, it aggressively discharges the storage, particularly in the middle of the two days. The power block is scheduled to operate at 56.5% capacity overnight in Heuristic-1, which 36.5% higher than SMILP for the same period. Consequently, the storage becomes empty which enforces a forced shutdown at early morning of the second day in many scenarios, see day 2 scenario 4. The overall simulation results show that dispatching under Heuristic-1 increases the average number forced shutdown events in the power block, which negatively influencing on increasing the degradation of subsystems as well as downgrading the profit of dispatching as missing the generation during favourable peak prices. This will be discussed further in the following subsection.

**Figure 6: The receiver's plan and control variables under Heurisitc-1**

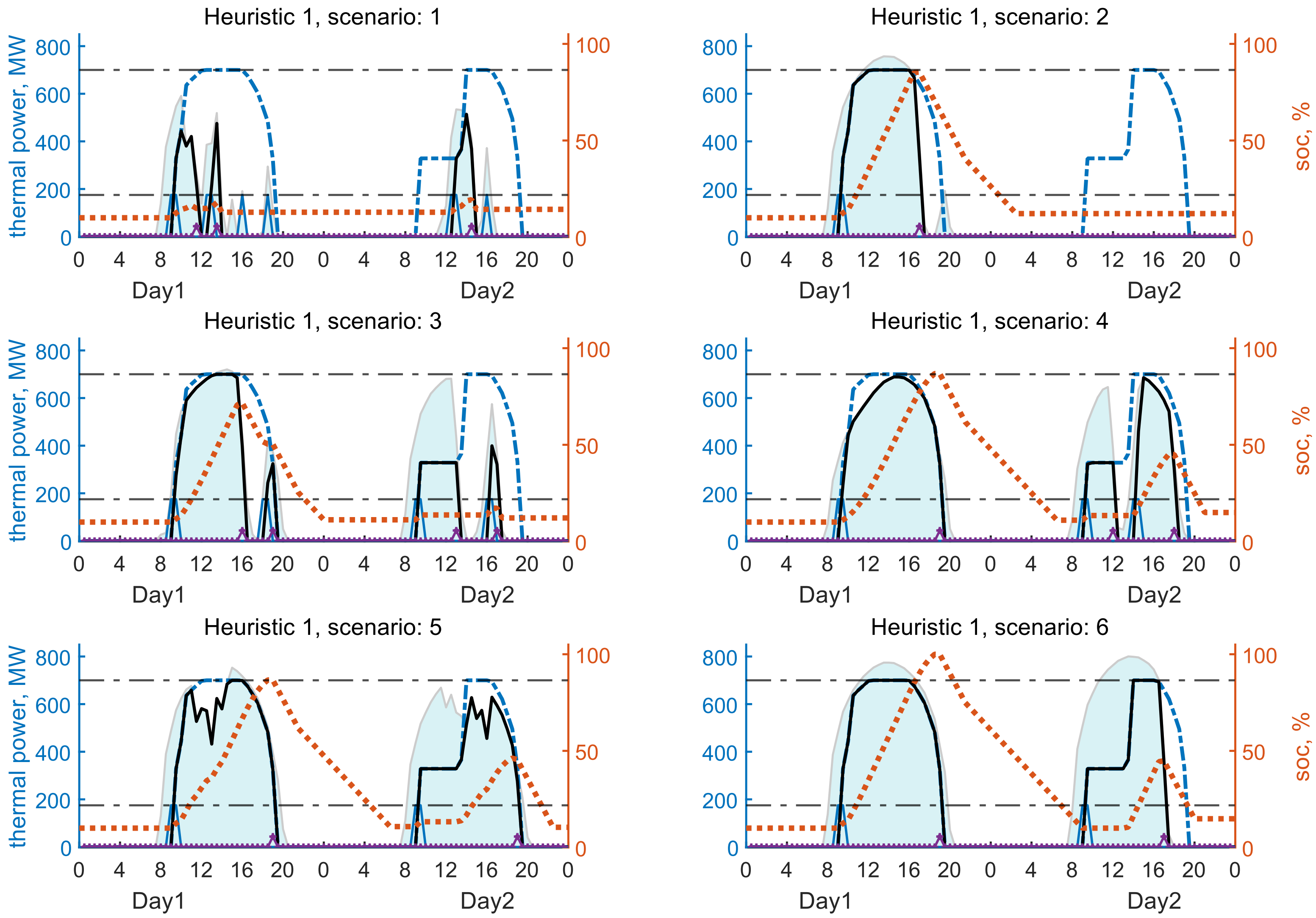

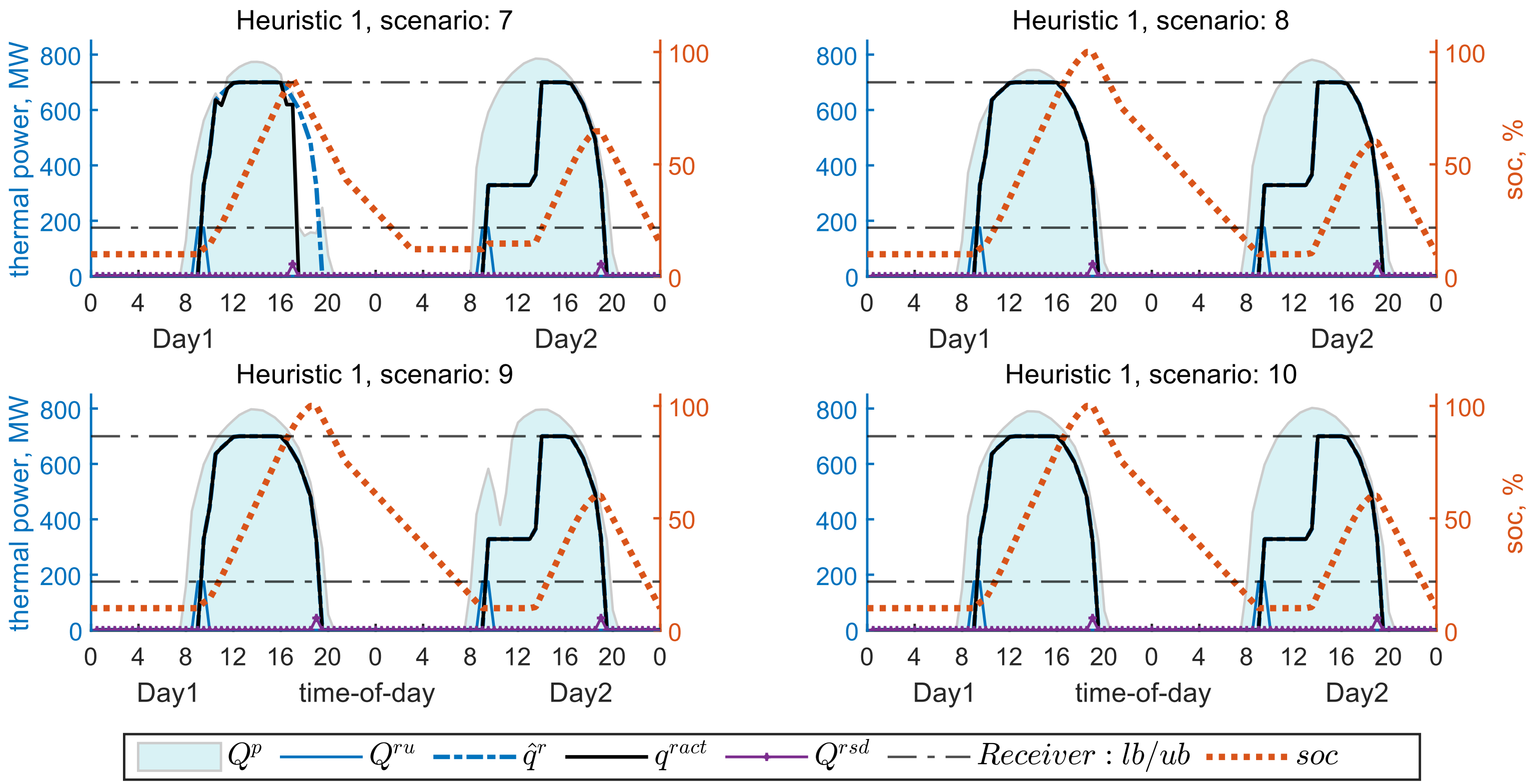
Heuristic 1, scenario: 7
Heuristic 1, scenario: 8
Heuristic 1, scenario: 9
Heuristic 1, scenario: 10
thermal power, MW
soc, %
0 4 8 12 16 20 0 4 8 12 16 20 0
Day1
Day2
time-of-day
$Q^p$ $Q^{ru}$ $\hat{q}^r$ $q^{ract}$ $Q^{rsd}$ $Receiver: lb/ub$ $soc$

**Figure 7: The power block (PB) plan and control variables under Heuristic-1.**

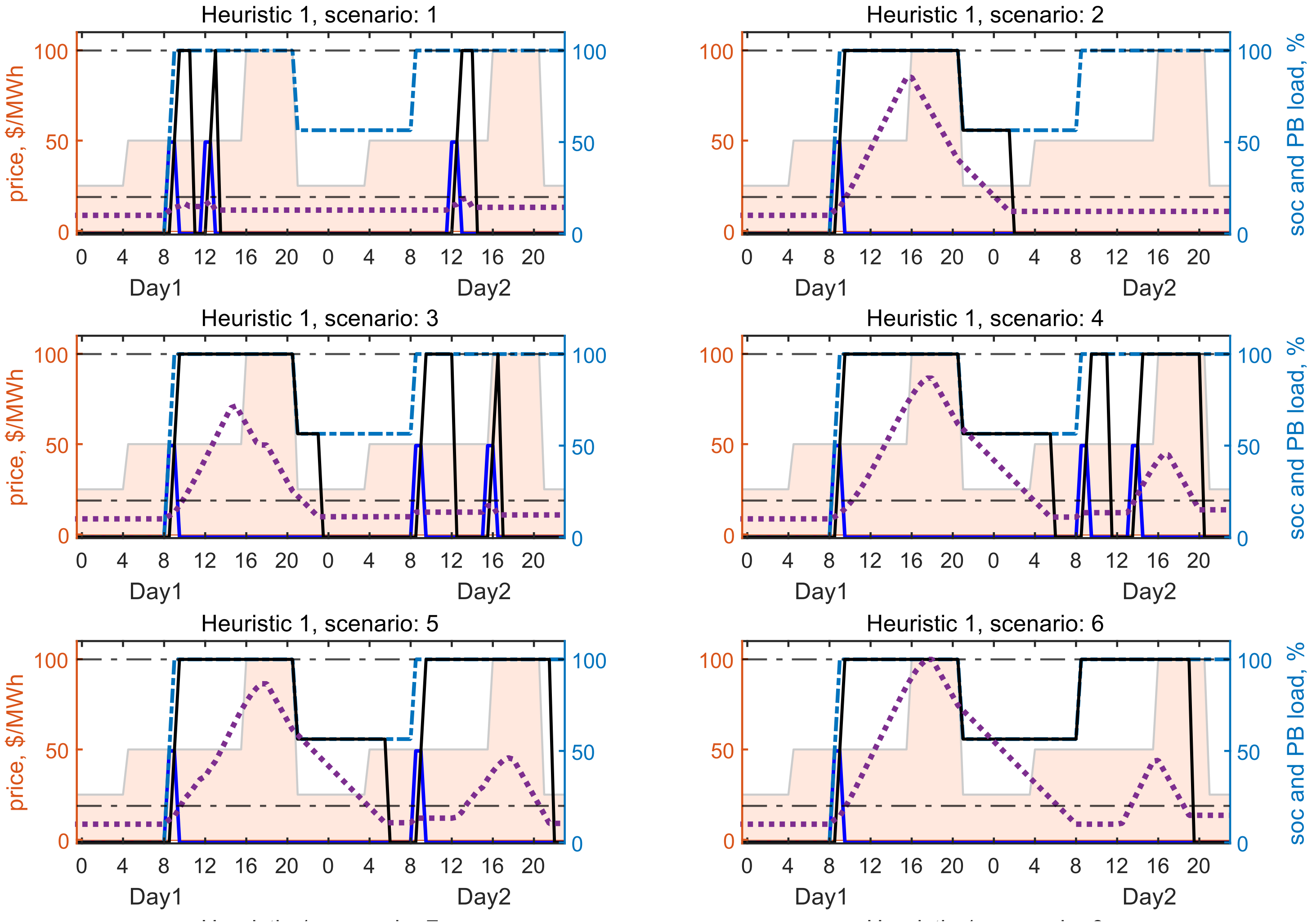

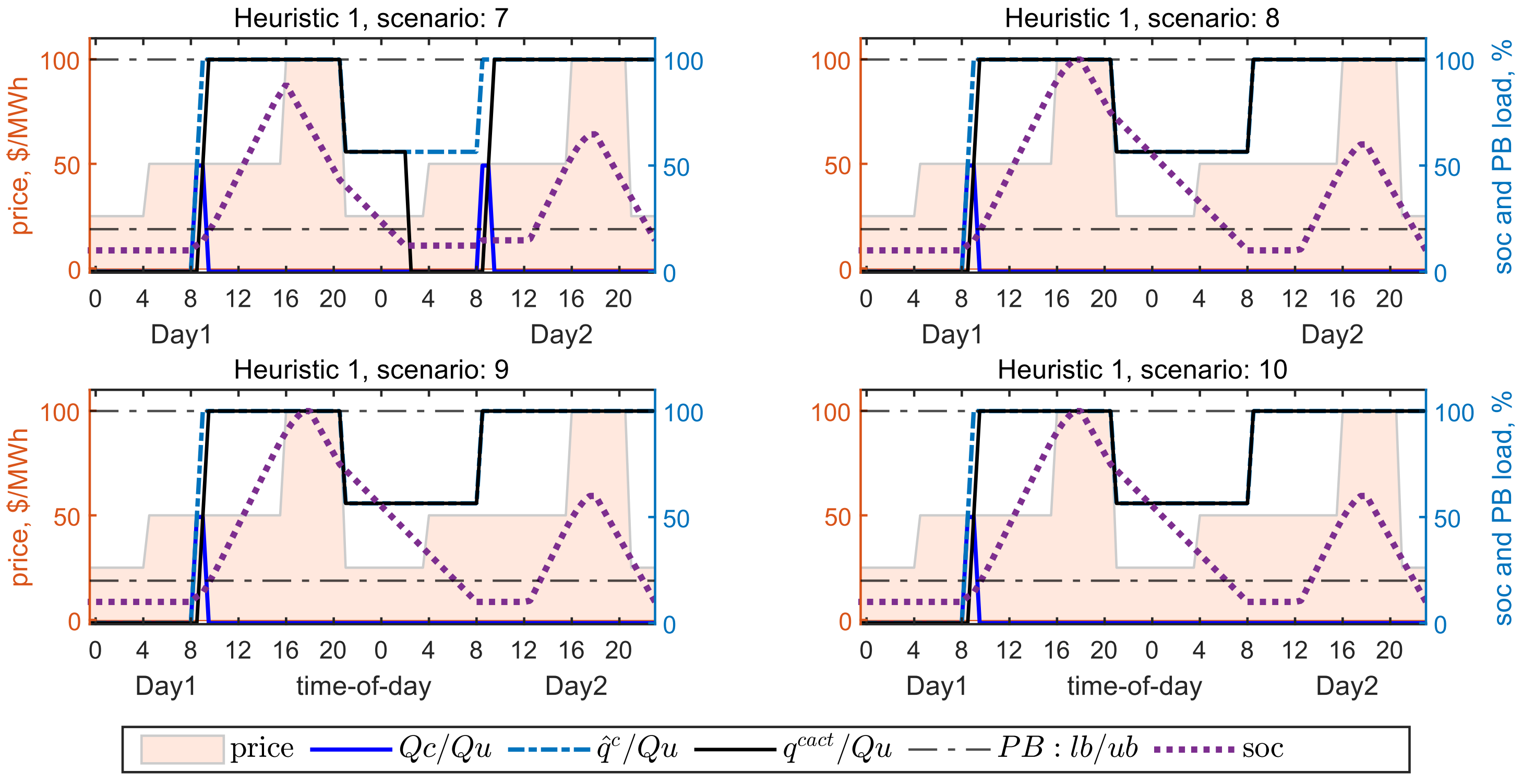
Heuristic 1, scenario: 7
Heuristic 1, scenario: 8
Heuristic 1, scenario: 9
Heuristic 1, scenario: 10
price, $/MWh
soc and PB load, %
0
50
100
0 4 8 12 16 20 0 4 8 12 16 20
Day1
Day2
time-of-day
price
$Qc/Qu$
$\hat{q}^c/Qu$
$q^{cact}/Qu$
$PB : lb/ub$
soc

0

### 6.1.3 Profit and cost evaluation for illustrative example

Table 3 reports the numerical results from dispatching under SMILP, PK, and Heuristic-1 presented in this illustrative example. As reported, the average heliostat field generation is approximately 12,430 MWht, and the average convective and radiative losses from the receiver is about 765 MWht across the scenario space. This results the potential thermal power available for the receiver to charge the storage being 11,665 MWh for all three DPs.

The simulation results show that as Heuristic-1 overestimates the solar irradiation, it generates an aggressive DP to operate the receiver. This has led the receiver generation in Heurisic-1 is averaged at 8,550 MWht. Conversely, SMILP is conservative in charging the storage, as it does not intend to impose a receiver forced shutdown due to overcharging the storage. This has resulted the average of generation curtailed increasing by 4.6% in SMILP, and startup and shutdown are slightly less than the counterparts in Heuristic-1.

The simulation results indicate that Dispatch Weighted Average (DWA) price [61] – which measures the value of 1 KWhe dispatched to the grid – reaches $63.4/MWhe for the case of SMILP. This is nearly equal to PK but significantly improved compared to Heuristic-1. This means although the average electricity generation dispatched to the grid in the case of SMILP is 5.0% less than Heuristic-1, dispatching using SMILP allows the majority of electricity generation at high prices. In addition, one major contributing factor driving the revenue is the power block forced shutdowns, which increase by 50% in Heuristic-1 compared to SMILP.

All that said, the estimate of expected profit in the case of SMILP is about $81,710, representing a 14.3% reduction compared to PK but a 15.2% improvement compared to Heuristic-1. In summary, although SMILP, with comprehensive consideration of the scenario space, generates a conservative plan to use the storage, it yields DWA and expected profit closer to the ideal PK than the Heuristic.

**Table 3: SMILP and benchmarks with evaluated on $\mathbb{W}_{N_s=10}^{SMILP}$, average across all scenarios**

| Simulation parameters | Unit | PK | SMILP | Heuristic-1 |
|---|---|---|---|---|
| Heliostat field generation | MWht | 12,430 | 12,430 | 12,430 |
| Thermal losses from Receiver[1] | MWht | 765 | 765 | 765 |
| Potential thermal power | MWht | 11,665 | 11,665 | 11,665 |
| Receiver generation | MWht | 8,550 | 8,155 | 8,555 |
| Receiver startups (shutdowns) | - | 1.8 (1.8) | 2.4 (2.2) | 2.6 (2.3) |
| Dispatch weighted average | $/MWhe | 63.8 | 63.4 | 60.9 |
| Electricity generation sales to grid | MWhe | 2,820 | 2,630 | 2,770 |
| Power block's startups (shutdowns) | - | 1.1 (0.7) | 1.2 (0.6) | 1.8 (1.4) |
| Profit from dispatching | $ | 95,400 | 81,710 | 70,915 |

[1] convective and radiative losses

## 6.2 The evaluation of SMILP and benchmarks

In this section, SMILP is provided a larger scenario space with $N_s = 14$, i.e., $\mathbb{W}_{N_s=14}^{SMILP}$. A simulation model, previously developed in [62], is employed. The performance of SMILP and benchmarks is evaluated when DPs applied to the CST system which is under the influence of weather profiles within one of the following categories: 1) scenario space $\mathbb{W}_{N_s=14}^{SMILP}$, 2) sampling set $\mathbb{W}_{N=339}^{sampling}$, and 3 testing set $\mathbb{W}_{N=153}^{testing}$. Figure 8.a shows the operating load planned for the receiver under SMILP and heuristic benchmarks, and Figure 8.b shows those for the power block.

As evident in Figure 8.a, SMILP tends to initiate receiver startup with a slight delay compared to other plans on the first day and shifts the shutdown to earlier hours to mitigate the risk of unplanned shutdowns at the end of each day. All DPs, except Heuristic-3, allow the receiver to operate at its maximum capacity on the first day. However, Heuristic-3 adapts to intraday variations in solar irradiation within the median weather profile, resulting in dynamic receiver operation. On the second day, both SMILP and Heuristic-2 accelerate the receiver to higher capacities in the early hours, while Heuristic-1 defers full-capacity receiver operation to later hours due to its excessive optimism regarding solar irradiation.

The exploration of Figure 8.b shows that the key distinctions among different DPs for the power block operation are evident in the timing of planned startup and the operating load during night-time hours. It is noticeable that SMILP's startup occurs earlier than Heuristic-1 and later than other heuristics. It is also witnessed that the power block is scheduled to operate at full capacity during the day, with the exception of night-time hours. Heuristic-1, which exhibits excessive confidence in solar irradiation, plans to maintain the power block at 53.9% of its full capacity. Heuristic-3, responding to intraday variations in the given solar profile, schedules the power block to operate at 36.4% of its full capacity. SMILP and Heuristic-2, being two conservative plans, plans to operate the power block at a load at 48.8% and 47.56% of its full capacity to save storage for dispatching during peak prices. In the following, the simulation results from implementation of DPs will be evaluated on three categories defined above.

**Figure 8: The planned load on critical subsystems: a) receiver, and b) power block**

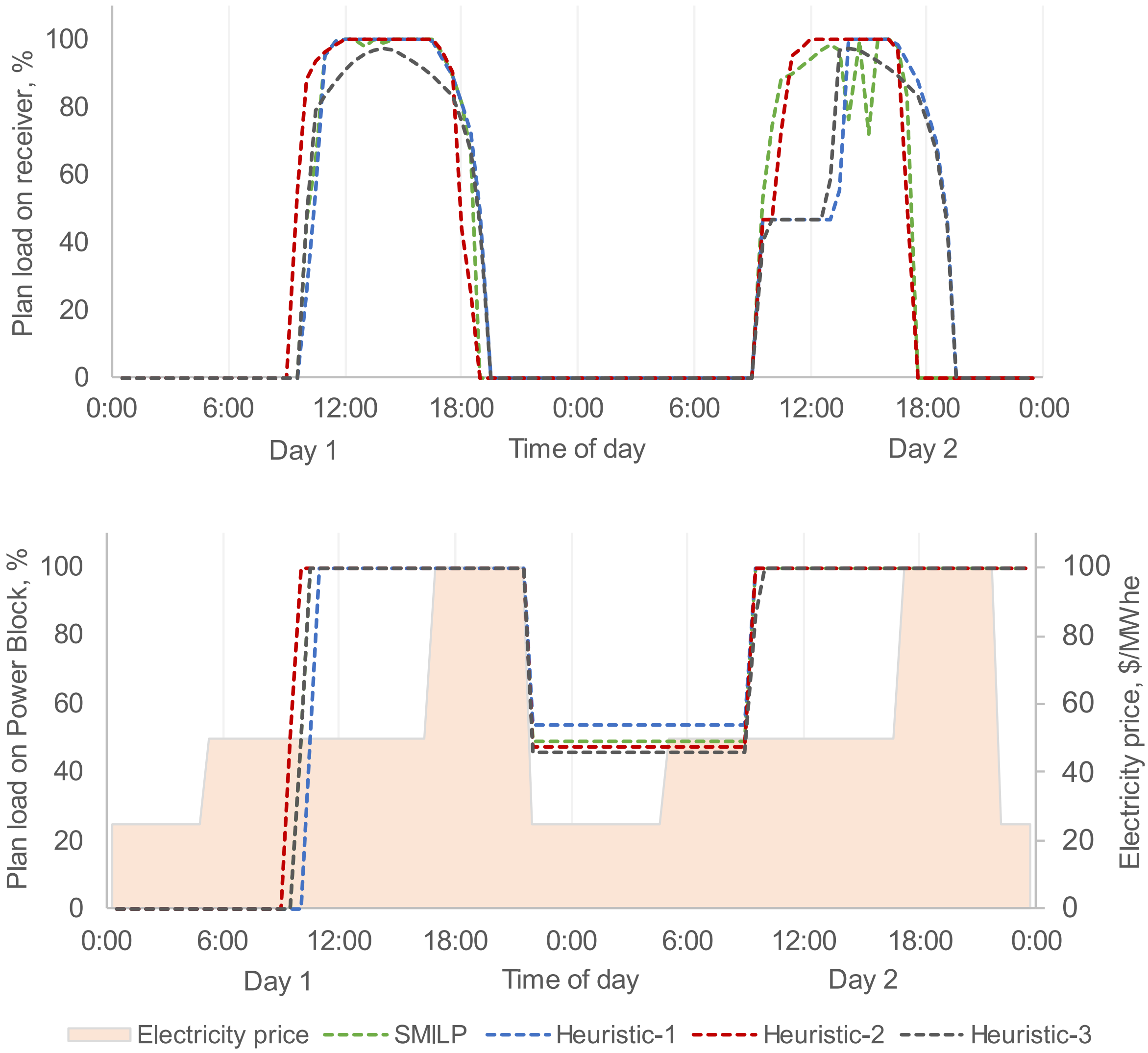

### 6.2.1 Evaluation of expected profit in category one – scenario set

In this section, DPs apply to the system while interacting with weather profiles within $\mathbb{W}_{N_s=14}^{SMILP}$. Table 4 reports the expected attainable revenue, incurred costs, and dispatching profit corresponding to each DP.

The first DP is PK which assumes weather trajectories are perfectly known. With PK, the expected revenue and profit are estimated at $181,690 and $96,390, respectively. As is evident, there is a considerably gap between the expected values obtained from PK with those from imperfect plans, which shows the importance of considering uncertainty in weather variables. SMILP and Heuristic-2 show closely aligned expected revenue – both achieving nearly 93.4% of ideal PK profit. The employment of other two heuristics drives the expected revenue down to ~90 % of that estimated via PK. Heuristic-1, for instance, operates the power block at a higher load overnight and postpones charging the storage to the late hours in the second day, which consequently leads to empty storage, and thus missing the revenue generation during evening high-price intervals in the second day.

Empty storage increases unplanned power block's shutdowns, while extended receiver operation raises the risk of forced receiver's shutdowns, both contributing to total degradation costs. SMILP incurs the lowest degradation costs among imperfect DPs due to its optimized solution minimizing forced shutdowns in critical subsystems. As reported in Table 4, SMILP drives down the operating cost in the power block by 6.3% compared to Heuristic-1, 4.4% compared to Heuristic-2, and 11.3% compared to Heuristic-3. The exploration of results also shows that SMILP decreases power block shutdown costs by 9.5% compared to Heuristic-1 and 2, and 17.4% compared to Heuristic-3. These increases in shutdown costs in Heuristic benchmarks are mainly due to more frequent forced shutdowns.

In summary, SMILP, with 88.7% of the optimal expected profit estimated via PK, outperforms the competitive heuristic benchmark, Heuristic-2, by 4.5%. This result aligns with the expectations because SMILP solution is derived through solving the proposed optimization model provided with $\mathbb{W}^{SMILP}_{N_s=14}$. Therefore, it is reasonable that SMILP performs better than the heuristics when the plant encounters weather profiles within $\mathbb{W}^{SMILP}_{N_s=14}$.

**Table 4: Expected revenue, degradation costs, and profit for SMILP and benchmarks evaluated on i.e., $W^{SMILP}_{N_s=14}$; Negative values (i.e., costs) are reported in ().**

| Profit components | PK | SMILP | Heuristic-1 | Heuristic-2 | Heuristic-3 |
|---|---|---|---|---|---|
| Revenue | $181,690 | $170,670 | $164,660 | $169,730 | $163,580 |
| Purchase cost* | ($10,920) | ($9,790) | ($10,470) | ($9,670) | ($10,310) |
| Receiver | | | | | |
| Operating cost | ($32,030) | ($30,700) | ($30,160) | ($30,900) | ($29,840) |
| Shutdown cost | ($13,500) | ($14,500) | ($15,500) | ($16,000) | ($15,500) |
| Power block | | | | | |
| Operating cost | ($21,850) | ($22,750) | ($24,290) | ($23,800) | ($25,660) |
| Shutdown cost | ($7,000) | ($7,400) | ($8,180) | ($8,170) | ($8,960) |
| Profit | $96,390 | $85,530 | $76,060 | $81,190 | $73,310 |

* cost of purchasing electricity from the grid to run the auxiliaries

### 6.2.2 Evaluation of expected profit in category two – sampling set

The same evaluation as in the previous section is conducted here, but the optimized plans are tested against weather profiles within $\mathbb{W}_{N=339}^{sampling}$.Figure 9 shows the average of revenue and incurred costs (i.e., purchase electricity cost, receiver and power block degradation costs) for different DP, and the relative difference with the PK counterparts.

As with category one, applying PK, with perfect insight assumption, leads to overestimated revenue and profit, along with underestimated degradation costs. The average revenue generation from applying ideal PK on sampling set is $176,460, while the expected purchase cost of $10,810, expected receiver and power block degradation costs of $44,390 and $24,640, respectively.

Among imperfect plans, SMILP captures up to 93.6% and 76.3% of the ideal expected revenue and profit, respectively. As seen in Figure 9, SMILP results an increase the total receiver and power block degradation costs by 7.9% and 37.0% relative to PK, respectively. On the contrary, Heuristic-2 performs the best, achieving 94.8% of the expected revenue and 78% of the expected profit. This is reasonable given the nature of Heuristic-2, which explores the DP driving the highest expected profit when applied to the sampling set. Additionally, the conservative approach to preserving storage is a key factor in reducing expected revenue in these two plans to contend with weather uncertainties.

Table 5 reports the median, and 2.5% and 97.5% percentile (pct) for attainable revenue, incurred costs, and the profit obtained from electricity sales from applying different DPs. It can be observed a consistent pattern across all DPs, with the median exceeding the mean in all distributions, resulting in left-skewed distributions – e.g., the median profit in the case of PK is estimated at $119,320, while the mean is estimated at $96,620. This skew is primarily due to the dominance of clear sky profiles over days with poor solar resources in the sampling set.

As seen in Table 5, the profit distributions exhibit a wide range in all DPs. The significant distinction is that PK ensures that profit does not drop below zero, thanks to its perfect insight that prevents the plant from starting up and incurring degradation costs on poor/cloudy days. In contrast, plants with imperfect DPs may experience profit losses in cloudy days. The results indicate a higher likelihood of reduced profit loss when employing SMILP and Heuristic-2 compared to the other two heuristics. The simulation results show not only the profit loss in SMILP and Heuristic-2 is smaller than other heuristics (i.e., see 2.5% pct), but also the expected profit in these two surpasses the other heuristics in clear-sky days (i.e., see 97.5% pct).

Finally, a t-test at a 5% significance level assesses the similarity of profit distributions obtained from the employment of different DPs. Results indicate significant differences between the profit distribution in imperfect DPs and PK (e.g., the p-value for SMILP vs. PK reaches 6.1576e-10). In contrast, profit distributions among imperfect plans do not significantly differ (e.g., p-value of 0.6891 for SMILP-Heuristic-2). However, computing time remains a challenge for SMILP optimization, taking 12 hours to solve optimization program given $\mathbb{W}_{N_s=14}^{SMILP}$, compared to 3 hours in Heuristic-2. In the next section, DPs are assessed when the plant encounters novel weather profiles.

**Figure 9: The average revenue and different degradation costs from applying DPs on sampling set; and % of relative difference compared to PK;**

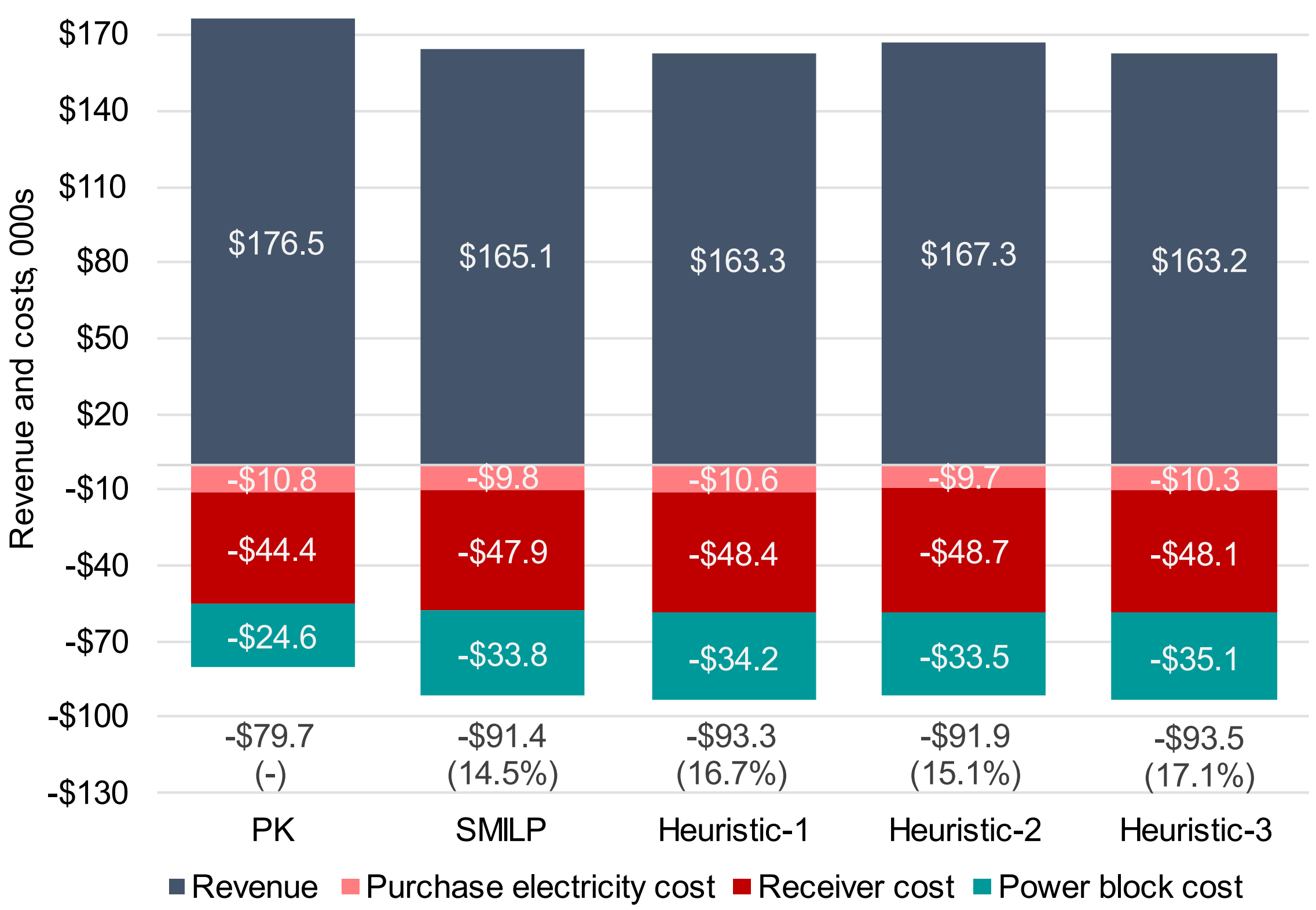

**Table 5: The evaluation of DPs on sampling set; negative values (costs) are in ()**

| DPs | Statistics | Revenue | Purchase | Receiver cost | Power Block cost | Profit |
|---|---|---|---|---|---|---|
| PK | 2.5% pct | $0 | $0 | $0 | $0 | $0 |
| | Median | $213,030 | ($13,050) | ($53,250) | ($24,600) | $119,320 |
| | 97.5% pct | $222,520 | ($14,450) | ($56,540) | ($40,470) | $128,560 |
| SMILP | 2.5% pct | $0.0 | ($180) | ($7,000) | $0 | ($56,170) |
| | Median | $199,410 | ($11,630) | ($55,000) | ($36,390) | $95,260 |
| | 97.5% pct | $215,970 | ($12,620) | ($62,200) | ($63,650) | $124,500 |
| Heuristic-1 | 2.5% pct | $0.0 | ($180) | ($7,000) | $0 | ($69,350) |
| | Median | $195,900 | ($12,150) | ($55,200) | ($34,060) | $89,580 |
| | 97.5% pct | $215,810 | ($13,910) | ($64,720) | ($70,660) | ($123,540) |
| Heuristic-2 | 2.5% pct | $0.0 | ($180) | ($7,000) | $0 | ($56,000) |
| | Median | $200,870 | ($11,420) | ($55,180) | ($32,740) | $95,260 |
| | 97.5% pct | $218,540 | ($12,440) | ($62,200) | ($61,520) | $126,480 |
| Heuristic-3 | 2.5% pct | $0.0 | ($270) | ($7,000) | $0 | ($65,070) |
| | Median | $195,000 | ($11,730) | ($55,210) | ($36,180) | $89,400 |
| | 97.5% pct | $213,550 | ($13,230) | ($65,000) | ($66,280) | $121,020 |

### 6.2.3 Evaluation of expected profit in category three – testing set

In this section, the performance of the SMILP solution and benchmarks is examined when the plant interacts with novel weather profiles within $\mathbb{W}_{N=153}^{testing}$. These weather profiles were neither used in the sampling process nor in the DP optimization. Figure 9 shows the average of revenue and incurred costs for different DPs, and the relative difference with the PK counterparts.

Table 6 presents a summary of statistics for the simulation results in category three. The exploration of results shows that the ideal expected revenue and the ideal receiver degradation costs drop 10.5%, compared to the counterpart in category two, while the ideal power block degradation cost only 6.9%. These declines which consequently drive down the ideal expected profit by 11.1% is mainly because of a higher frequency of cloudy days in testing set than sampling set. This is indicated by the gap between the two ECDFs in Figure 2. The same reason also drives the gap between the mean and median in the profit distribution of the case of PK down by 9.4%, and further decreases are observed in imperfect dispatch schedules, e.g., 60% in SMILP. As with category two, Heuristic-2 outperforms the SMILP and the other two heuristics, but its gap with SMILP shrinks due to the nature of the testing set.

In comparison with category two, the 2.5% pct and 97.5% pct in the profit distribution of all dispatching plans are slightly wider in category three. Although the 97.5% pct of profit are nearly identical for both categories, the higher frequency of cloudy days leads the 2.5% pct to slightly decrease for testing set. The results of the t-test also indicate that the profit distribution for PK significantly differs from imperfect DPs (p-value < 2.95e-5), while SMILP and Heuristic-2 exhibit no significant difference, with a p-value of 0.9186. In summary, considering computation time, Heuristic-2 emerges as a viable solution for estimating expected profit from dispatching, serving as an efficient alternative to SMILP. In contrast, Heuristic-3, which relies on representative weather profiles, is not recommended due to its lower performance.

**Figure 10: The average revenue and different degradation costs from applying DPs on testing set; and % of relative profit difference compared to PK**

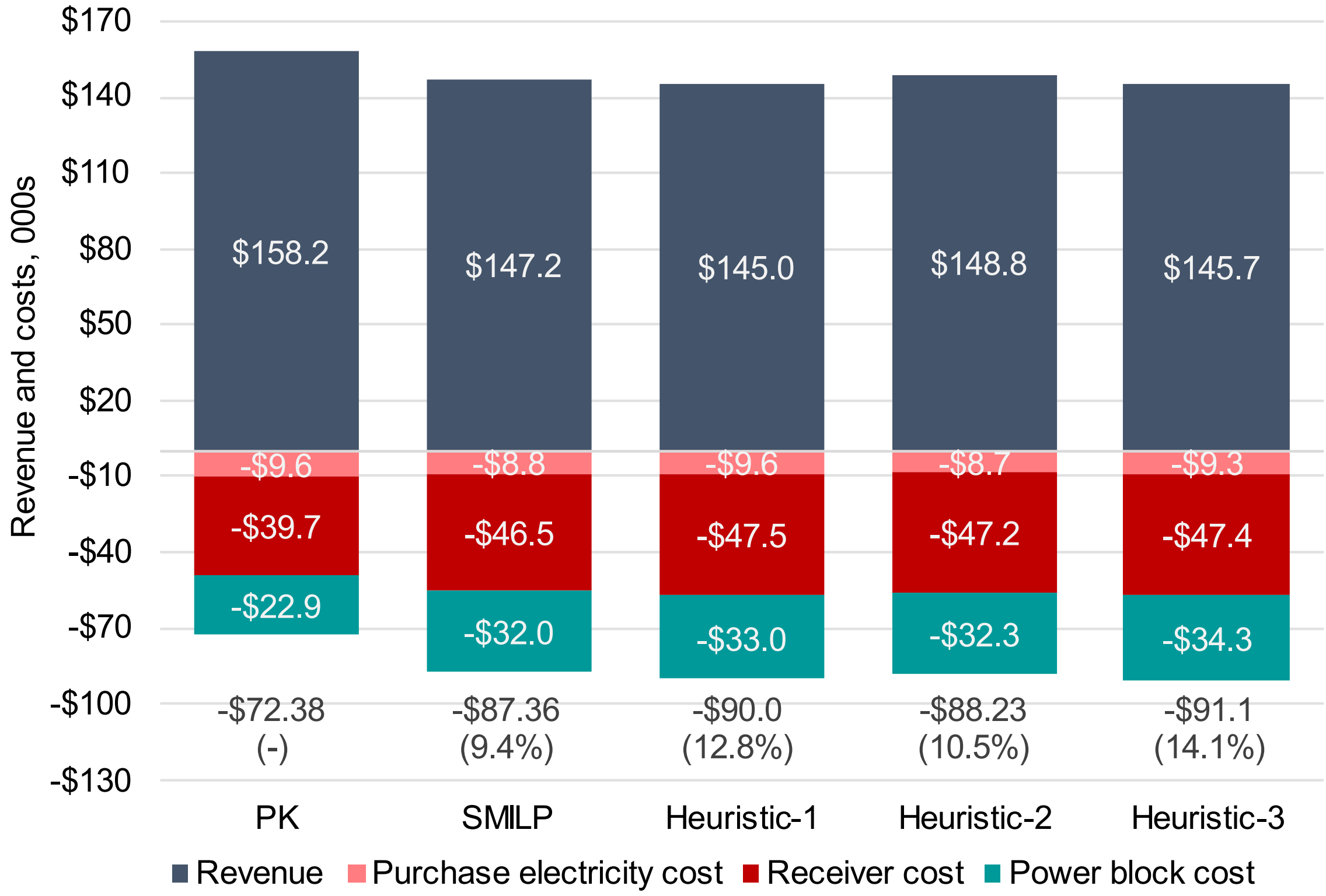

**Table 6: The evaluation of DPs on the testing set; negative values (i.e., costs) are reported in ()**

| DPs | Statistics | Revenue | Purchase | Receiver cost | Power Block cost | Profit |
|---|---|---|---|---|---|---|
| PK | 2.5% pct | $0 | $0 | $0 | $0 | $0 |
| | Median | $195,920 | ($11,120) | ($48,550) | ($24,610) | $106,660 |
| | 97.5% pct | $222,520 | ($14,480) | ($55,220) | ($38,250) | $128,500 |
| SMILP | 2.5% pct | $0 | ($30) | ($2,270) | $0 | ($59,590) |
| | Median | $172,640 | ($10,060) | ($54,590) | ($32,320) | $68,350 |
| | 97.5% pct | $215,970 | ($12,620) | ($62,130) | ($60,650) | $124,480 |
| Heuristic-1 | 2.5% pct | $0 | ($30) | ($2,270) | $0 | ($67,430) |
| | Median | $165,410 | ($10,650) | ($55,220) | ($31,790) | $62,300 |
| | 97.5% pct | $215,810 | ($13,920) | ($66,780) | ($64,520) | $123,450 |
| Heuristic-2 | 2.5% pct | $0 | ($30) | ($2,270) | $0 | ($69,640) |
| | Median | $174,800 | ($10,140) | ($54,610) | ($32,530) | $67,820 |
| | 97.5% pct | $218,540 | ($12,440) | ($62,450) | ($61,330) | $126,460 |
| Heuristic-3 | 2.5% pct | $0 | ($30) | ($2,270) | $0 | ($67,080) |
| | Median | $175,000 | ($10,770) | ($55,220) | ($36,110) | $66,720 |
| | 97.5% pct | $213,560 | ($13,240) | ($66,300) | ($63,600) | $121,020 |

### 6.2.4 Sensitivity on the scenario space in Heuristic-2

As shown in the previous section, the expected profit obtained from applying SMILP and Heuristic-2 on the testing set are quasi-identical, meaning that Heuristic-2 can be used in place of SMILP in estimation of expected profit. In this section, a sensitivity analysis is conducted where scenario space provided to Heuristic-2 is a subset of sampling set such that $N_{H2} = 50, 100, \dots, 300$, where $N_{H2}$ is number of recent weather profiles in the subset. The DP is applied to the testing set and the calculated expected profit compared with SMILP and Heuristic-3.

Figure 11 shows expected profit when different DPs applied to testing set (left axis) and the computing time to find the optimized DP via Heuristic-2 (right axis). As is evident, SMILP outperforms Heuristic-2 unless the $N_{H2}$ exceeds from 150 profiles. The expected profit in the case of SMILP is about 9% and 6% higher than Heuristic-2 with $N_{H2} = 90$ and $140$ weather profiles, respectively. The performance becomes quasi-identical between SMILP and Heuristic-2 when $N_{H2}$ varies in the range of 150 to 220, whereas computing time to find optimized DP in Heuristic-2 increases 83%. Even if the size of scenario space increases to 220 profiles, only 1% improvement in expected profit is achieved via Heuristic-2 compared to SMILP, but the computing time is approximately two times higher (i.e., increasing one hour).

In summary, it does not necessarily need to employ all the historical weather profile in selecting the optimized DP in Heuristic-2 to achieve performance close to SMILP, particularly when considering the computing time. In the case of having small or unreliable sampling set, the deployment of Heuristics increases the risk of losing profit, whereas SMILP can be a more effective approach allowing to achieve higher expected profit when applying for dispatching under novel weather profiles.

**Figure 11: Sensitivity analysis on the size of scenario space provided to the Heuristic-2, expected profit (right axis) and computing time to find optimized DP (left axis)**

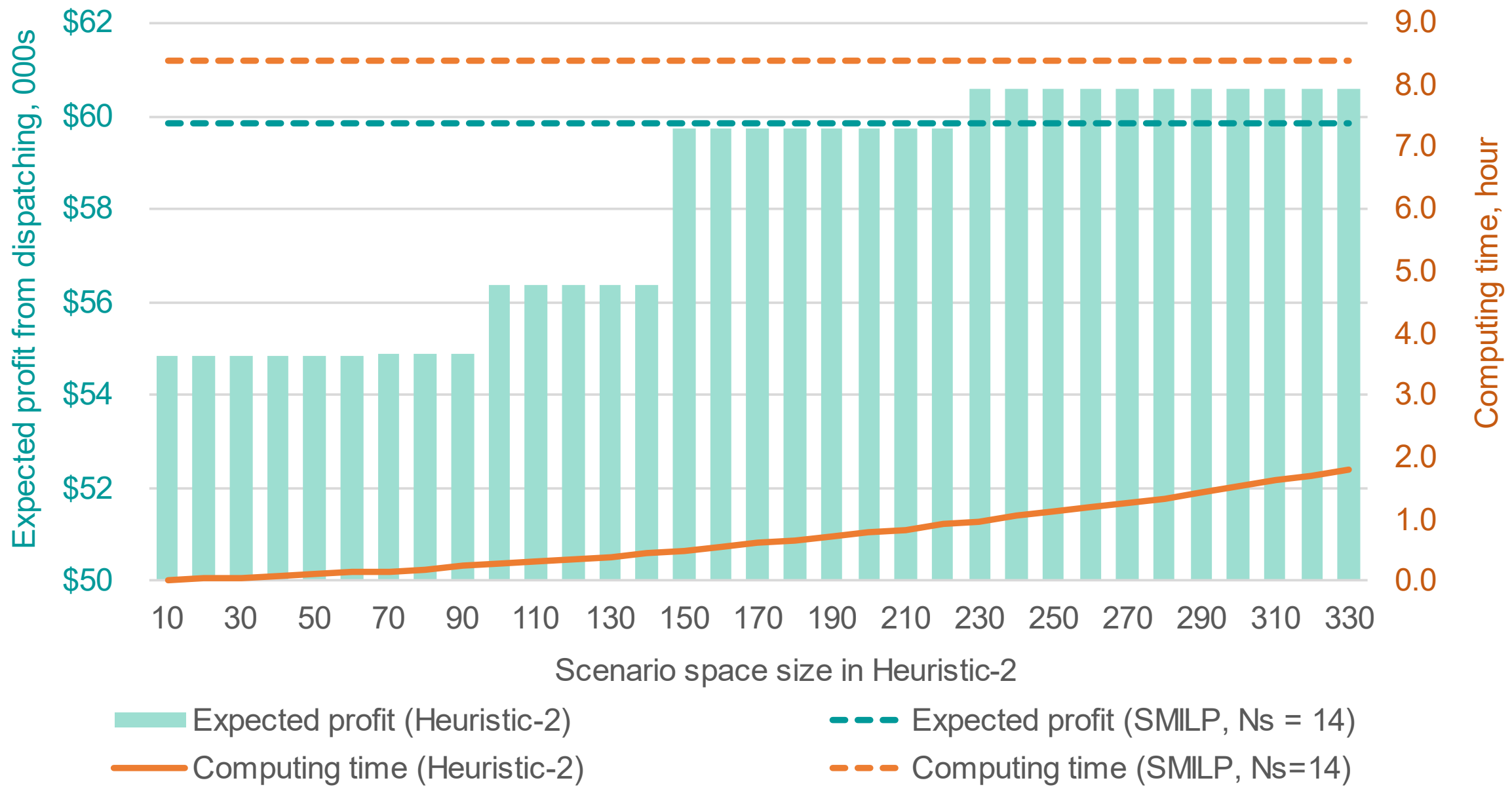

## 7 Conclusion

The key intention of this study is to develop a novel Stochastic Mixed Integer Linear Program (SMILP) for dispatch optimization in Concentrated Solar Thermal (CST) plant integrated a Thermal Energy Storage (TES). The proposed SMILP model maximizes the Sample Average Approximation (SAA) of expected profit considering a scenario space for weather variables. Unlike chanced constrained optimization, the proposed SMILP avoids probabilistic constraints, and the interaction between physical constraints and uncertainty are expressed through linear logical constraints. This approach allows the solution generated via SMILP to be feasible for all scenarios, and SAA estimates the expected profit more accurately.

In a case study, the SMILP solution applies to a hypothetical 115 MW CST plant with a 10-hour TES operating under a two-tier price agreement. A previously developed simulation model [24] is employed to evaluate the performance of SMILP against one idealized perfect knowledge (PK) strategy and three heuristic approaches. The historical weather data are divided into sampling set and testing set, where the first set is used to generate scenario for SMILP, while testing set is maintained as novel weather profit to test the performance of Dispatch Plans (DPs). SMILP and benchmarks are evaluated under three categories of weather variables: scenario space, sampling set, and testing set.

The exploration of simulation results shows that SMILP outperforms heuristic approaches when the plant encounters weather profiles within the scenario space, achieving 88.7% of the ideal expected profit, which exceeds the most competitive Heuristic (i.e., Heuristic 2) by at least 4.5%. However, when DPs are applied to sampling set, Heuristic-2 performs better reaching 78% of the ideal profit estimated, compared to SMILP's 76.3%. The frequency of cloudy days is higher in the testing set leads the gap between SMILP and Heuristic-2 to be narrowed, but Heuristic-2 still slightly outperforming SMILP by 1.2%. The results of t-tests at a 5% significance level indicate that profit distributions obtained from SMILP and Heuristic-2 were not significantly different, in contrast to PK dispatching, which exhibits significant differences across all categories. These findings highlight the importance of accounting for solar irradiation uncertainty in profit evaluations while underscoring Heuristic-2 as a practical choice for decision-makers in CST plant operation and planning, considering the constraints of time and data availability.

Finally, a sensitivity analysis is also conducted on the number of historical weather profiles that must be provided to Heuristic-2 to achieve the SMILP output when applied to testing set. The results reveal that the expected profit estimated via Heuristic-2 can be 6% to 9% less than the counterpart in SMILP unless the size of scenario space provided to Heuristic includes more recent weather profiles. The results show that Heuristic-2 can achieve performance close to SMILP with the half of historical profiles within a manageable timeframe. Future research could explore integrating optimized DPs into a rolling horizon control framework and delve into the impact of volatile electricity prices on dispatch planning for CST technology.

## Acknowledgement

The authors acknowledge the support of the Australian Government for this study, through the Australian Renewable Energy Agency (ARENA) and within the framework of the Australian Solar Thermal Research Institute (ASTRI).

# Appendix A.1 Nomenclature

**Table A.1: sets and parameters**

| | |
|---|---|
| Sets at time $k$ | |
| $\eta_k^c$ | Normalized condenser parasitic $(-)$ |
| $\eta_k^{opt}$ | Optical efficiency $(-)$ |
| $c_k$ | Heliostat availability $(-)$ |
| $\rho_k$ | Mirror reflectance and soiling coefficient $(-)$ |
| $\lambda^k$ | Discount rate (e.g.,, $0.98^k$) |
| Sets at time $k$ scenario $s$ | |
| $f_{k,s}^R$ | Revenue from dispatching ($) |
| $f_{k,s}^{aux}$ | Cost of purchasing electricity from the grid to run auxiliaries ($) |
| $f_{k,s}^{om}$ | Degradation costs associated with key subsystems ($) |
| $Q_{k,s}^{helio}$ | Thermal power generated by heliostat field $(kWt)$ |
| $Q_{k,s}^{rad}$ | Radiative loss from receiver to ambient $(kWt)$ |
| $Q_{k,s}^{conv}$ | Convective loss from receiver to ambient $(kWt)$ |
| $Q_{k,s}^{dump}$ | Thermal power dumped from receiver to ambient $(kWt)$ |
| $Q_k^p$ | Potential thermal power the receiver to use $(kWt)$ |
| $Q_{k,s}^{avail}$ | Available thermal power available for receiver to use post startup/shutdown $(kWt)$ |
| $Q_{k,s}^{avail,gen}$ | Available thermal power when the receiver is in generating mode $(kWt)$ |
| $Q_{k,s}^{dump}$ | Thermal power dumped to ambient from receiver at time $k$ in scenario $s$ $(kWt)$ |
| $p_{k,s}$ | Electricity price $(\$/kWhe)$ |
| $DNI_{k,s}$ | Direct normal irradiation $(kW/m^2)$ |
| $T_{k,s}^{amb}$ | Ambient temperature $(K)$ |
| $v_{k,s}$ | Wind speed $(m/s)$ |
| Degradation cost | |
| $C^{rec}$ | For each $kWh$ thermal power generation in receiver $(\$/kWht)$ |
| $C^c$ | For each $kWh$ electricity generation in power block $(\$/kWhe)$ |
| $C^{rsup}$ | For each receiver cold start-up event $(\$/start)$ |
| $C^{csup}$ | For each power block cold start-up event $(\$/start)$ |
| $C^{\delta W}$ | For ramping up/down the power block $(\$/\Delta kW_e)$ |
| Heliostat field and receiver parameters | |
| $A^{helio}$ | Heliostat reflective area $(m^2)$ |
| $E^r$ | Thermal energy consumed in the receiver to complete start-up process $(kWht)$ |
| $E^{hs}$ | Electrical parasitic loss in heliostat field during start-up or shutdown $(kWhe)$ |
| $N^{helio}$ | Total number of heliostats $(-)$ |
| $L^{r}$ | Electrical parasitic loss in receiver pumping system $(kWe/kWt)$ |
| $Q^{pipe}$ | Thermal power loss in the piping system $(kWt)$ |
| $Q^{ru}$ | Receiver allowable thermal power per time step in start-up event $(kWt)$ |
| $Q^{rl}$ | Receiver minimum allowable operating load $(kWt)$ |
| $Q^{rlim}$ | Receiver maximum allowable operating load $(kWt)$ |
| $Q^{rsd}$ | Thermal power to shut down the receiver $(kWt)$ |
| $W^{h}$ | Electrical parasitic loss in heliostat tracking system $(kWe)$ |
| Power block parameters | |
| $E^{c}$ | Thermal energy consumed in power block to complete start-up process $(kWht)$ |
| $\eta^{p}$ | Slope of linear approximation of power block performance curve $(kWe/kWt)$ |
| $L^{c}$ | Electrical parasitic loss in power block pumping system $(kWe/kWt)$ |
| $Q^{c}$ | Allowable thermal power per time step to start up the power block $(kWt)$ |
| $Q^{l}$ | Minimum allowable inlet power to the power block $(kWt)$ |
| $Q^{u}$ | Maximum allowable inlet power to the power block $(kWt)$ |
| $W^{l}$ | Minimum output from power block $(kWe)$ |
| $W^{u}$ | Maximum output from power block $(kWe)$ |
| Thermal storage parameter | |
| $E^{u}$ | Thermal energy storage capacity $(kWht)$ |
| $SOC_{min}$ | Minimum state of charge in storage $(-)$ |
| Miscellaneous parameters | |
| $\alpha$ | Conversion factor between dimensionless and monitory values |
| $\Delta t$ | Time step duration $(hr)$ |
| $\mathbb{M}$ | A very large positive number $(-)$ |
| $\epsilon$ | A very small positive number $(-)$ |
| $s$ | Scenario index $(-)$ |
| $N_{s}$ | Number of scenarios in scenario space $(-)$ |
| $k$ | Time index for future time instance $(-)$ |
| $K$ | Optimization horizon $(-)$ |

**Table A.2: continuous and binary decision variables**

| | |
|---|---|
| Continues plan variables, at time $k$ | |
| $\hat{q}_k^r$ | Plan storage charging rate $(kWt)$ |
| $\hat{q}_k^c$ | Plan storage discharge rate $(kWt)$ |
| Continuous control variables, at time $k$ in scenario $s$ | |
| $q_{k,s}^{ract}$ | Actual storage charting rate $(kWt)$ |
| $q_{k,s}^{cact}$ | Actual storage discharging rate $(kWt)$ |
| $e_{k,s}^{rsu}$ | Receiver thermal state during startup $(kWh)$ |
| $e_{k,s}^{csu}$ | Power block thermal state during startup $(kWht)$ |
| $\phi_{k,s}^r$ | Receiver slack variable during startup $(kWh)$ |
| $\phi_{k,s}^c$ | Power block slack variable during startup (kWh) |
| $\dot{w}_{k,s}$ | Power block rate of gross electricity generation $(kWe)$ |
| $\dot{w}_{k,s}^{\delta}$ | Power block rate of ramping up/down $(kWe)$ |
| $\dot{w}_{k,s}^{s}$ | Power block rate of generation dispatched to the market $(kWe)$ |
| $\dot{w}_{k,s}^{p}$ | Plant rate of electricity purchased from the grid to run axillaries $(kWe)$ |
| $\dot{w}_{k,s}^{net}$ | Plant rate of net electricity generation $(kWe)$ |
| $s_{k,s}$ | Storage reserved energy $(kWht)$ |
| $\varphi_{k,s}$ | Level of available reserve to storage lower bound $(kWht)$ |
| Binary plan variables | |
| $\hat{y}_k^r$ | = 1 if receiver is planned for generation at time $k$; otherwise, 0 |
| $\hat{y}_k^{rsup}$ | = 1 if receiver is planned for cold startup at time $k$; otherwise, 0 |
| $\hat{y}_k^{rsd}$ | = 1 if receiver is planned for shutdown at time $k$; otherwise, 0 |
| $\hat{y}_k^c$ | = 1 if power block is planned for generation mode at time $k$; otherwise, 0 |
| $\hat{y}_k^{csup}$ | = 1 if power block is planned for cold startup at time $k$; otherwise, 0 |
| $\hat{y}_k^{csd}$ | = 1 if power block is planned for shutdown at time $k$; otherwise, 0 |
| Binary control variables | |
| $\delta_{k,s}^{rsup}$ | = 1 if cold startup event occurs for the receiver at time $k$ in scenario $s$; otherwise, 0 |
| $\delta_{k,s}^{r}$ | = 1 if the receiver is in power generation mode at time $k$ in scenario $s$; otherwise, 0 |
| $\delta_{k,s}^{rsu}$ | = 1 if the receiver is during startup process at time $k$ in scenario $s$; otherwise, 0 |
| $\delta_{k,s}^{rsd}$ | = 1 if shutdown event occurs for the receiver at time $k$ in scenario $s$; otherwise, 0 |
| $\delta_{k,s}^{csup}$ | = 1 if cold startup event occurs for the power block at time $k$ in scenario $s$; otherwise, 0 |
| $\delta_{k,s}^{csu}$ | = 1 if the power block is during startup process at time $k$ in scenario $s$; otherwise, 0 |
| $\delta_{k,s}^{c}$ | = 1 if the power block is in power generation mode at time $k$ in scenario $s$; otherwise, 0 |
| $\delta_{k,s}^{csd}$ | = 1 if shutdown event occurs for the power block at time $k$ in scenario $s$; otherwise, 0 |
| Auxiliary binary variables | |
| $z_{k,s}^{(1)}$ | = 1 if $Q_{k,s}^{p} < Q^{rl}$; otherwise, 0 |
| $z_{k,s}^{(2)}$ | = 1 if $e_{k,s}^{rsu} = E^{r}$; otherwise, 0 |
| $z_{k,s}^{(3)}$ | = 1 if $Q_{k,s}^{avail} < Q^{rl}$; otherwise, 0 |
| $z_{k,s}^{(4)}$ | = 1 if $Q_{k,s}^{avail,gen} < Q^{rl}$; otherwise, 0 |
| $z_{k,s}^{(5)}$ | = 1 if $Q_{k,s}^{avail,gen} < \hat{q}_{k}^{r}$; otherwise, 0 |
| $z_{k,s}^{(6)}$ | = 1 if $e_{k,s}^{csu} = E^{c}$; otherwise, 0 |
| $z_{k,s}^{(7)}$ | = 1 if $\varphi_{k,s} < \Delta t \cdot Q^{c}$; otherwise, 0 |

# Appendix A.2: SMILP optimization

## A.2.1. Receiver logical planned constraints

Constraint (A.1) prevents the receiver from power generation behind daylight hours, where $t$ is the time of day, $t^{rise}$ is sunrise time, and $t^{set}$ is sunset time. Constraint (A.2) counts the planned cold startup events. Constraints (A.3) and (A.4) ensure the cold startup occurs in the first step and shutdown in the last step of the receiver's operation, respectively. Constraint (A.5) prevents the receiver from power generation at time index $k$ if a shutdown occurred in previous time index $k-1$.

$$\hat{y}_{k}^{r} = \begin{cases} 0 & k|\, t \notin \left(t^{rise}, t^{set}\right) \\ 0,1 & o.w. \end{cases} \qquad \forall k \qquad \text{(A.1)}$$

$$\hat{y}_{k}^{rsup} \geq \hat{y}_{k}^{r} - \hat{y}_{k-1}^{r} \qquad \forall k \qquad \text{(A.2)}$$

$$\hat{y}_{k}^{rsup} + \hat{y}_{k-1}^{r} \leq 1 \qquad \forall k \qquad \text{(A.3)}$$

$$\hat{y}_{k-1}^{rsd} \geq \hat{y}_{k-1}^{r} - \hat{y}_{k}^{r} \qquad \forall k \qquad \text{(A.4)}$$

$$y_{k-1}^{rsd} + y_{k}^{r} \leq 1 \qquad \forall k \qquad \text{(A.5)}$$

## A.2.2. Receiver startup

Two 0-1 switches $z_{k,s}^{(1)}$ and $z_{k,s}^{(2)}$ allow to define multi-statement logical constraints. The big-M method is employed to convert the multi-statement functions into linear constraints [63]:

$$-\mathbb{M}\cdot z_{k,s}^{(1)} \leq Q_{k,s}^{p} - Q_{k,s}^{rl} \leq \mathbb{M}\cdot\left(1 - z_{k,s}^{(1)}\right) - \epsilon \qquad \forall k, \forall s \in \mathbb{S} \qquad \text{(A.6)}$$

$$-\mathbb{M}\cdot\left(1 - z_{k,s}^{(2)}\right) \leq e_{k,s}^{rsu} - E^{r} \leq \mathbb{M}\cdot z_{k,s}^{(2)} - \epsilon \qquad \forall k, \forall s \in \mathbb{S} \qquad \text{(A.7)}$$

Constraints (A.3) to (A.7) translate Eq. (13) into a set of linear constraints:

$$\delta_{k,s}^{rsu} \geq \hat{y}_{k}^{r} - \left(\delta_{k-1,s}^{r} + z_{k,s}^{(1)} + z_{k-1,s}^{(2)}\right) \qquad \forall k, \forall s \in \mathbb{S} \qquad \text{(A.8)}$$

$$\delta_{k,s}^{rsu} \leq \hat{y}_{k}^{r} \qquad \forall k, \forall s \in \mathbb{S} \qquad \text{(A.9)}$$

$$\delta_{k,s}^{rsu} + z_{k,s}^{(1)} \leq 1 \qquad \forall k, \forall s \in \mathbb{S} \qquad \text{(A.10)}$$

$$\delta_{k,s}^{rsu} + z_{k-1,s}^{(2)} \leq 1 \qquad \forall k, \forall s \in \mathbb{S} \qquad \text{(A.11)}$$

$$\delta_{k,s}^{rsu} + \delta_{k-1,s}^{r} \leq 1 \qquad \forall k, \forall s \in \mathbb{S} \qquad \text{(A.12)}$$

### A.2.3. Receiver power generation

An auxiliary 0-1 switch $z_{k,s}^{(3)} = 1$ when total available thermal power is inadequate to satisfy the receiver's minimum generation limit:

$$\mathbb{M}\cdot\left(z_{k,s}^{(3)} - 1\right) \leq Q_{k,s}^{rl} - Q_{k,s}^{avail} \leq \mathbb{M}\cdot z_{k,s}^{(3)} - \epsilon \qquad \forall k, \forall s \in \mathbb{S} \qquad \text{(A.13)}$$

Constraints (A.14) to (A.17) translate the conditions in Eq. (19) into linear constraints:

$$\delta_{k,s}^{r} \geq \hat{y}_{k}^{r} + z_{k,s}^{(2)} + \delta_{k-1,s}^{r} - z_{k,s}^{(3)} - 1 \qquad \forall k \ \wedge \forall s \in \mathbb{S} \qquad \text{(A.14)}$$

$$\delta_{k,s}^{r} \leq \hat{y}_{k}^{r} \qquad \forall k \ \wedge \forall s \in \mathbb{S} \qquad \text{(A.15)}$$

$$\delta_{k,s}^{r} \leq z_{k,s}^{(2)} + \delta_{k-1,s}^{r} \qquad \forall k \ \wedge \forall s \in \mathbb{S} \qquad \text{(A.16)}$$

$$\delta_{k,s}^{r} + z_{k,s}^{(3)} \leq 1 \qquad \forall k \ \wedge \forall s \in \mathbb{S} \qquad \text{(A.17)}$$

Two 0-1 switches $z_{k,s}^{(4)}$ and $z_{k,s}^{(5)}$ are defined. Constraint (A.13) allows $z_{k,s}^{(4)} = 1$ if $Q_{k,s}^{avail,gen}$ falls below the minimum receiver generation limit at time $k$ in scenario $s$, zero otherwise. Constraint (A.14) ensures $z_{k,s}^{(5)} = 1$ when $Q_{k,s}^{avail,gen}$ is less than the plan charging rate $\hat{q}_{k}^{r}$, zero otherwise. Constraint (A.15) allows $q_{k,s}^{ract} = Q_{k,s}^{avail,gen}$ if $z_{k,s}^{(4)} = 0$ and $z_{k,s}^{(5)} = 1$, meaning $Q^{rl} \leq$ $Q_{k,s}^{avail,gen} < \hat{q}_k^r$. Constraint Constraint (A.15) drives $q_{k,s}^{ract} = \hat{q}_k^r$ if $z_{k,s}^{(5)} = 0$, meaning that $Q_{k,s}^{avail,gen} \geq \hat{q}_k^r$.

$$\mathbb{M} \cdot \left(z_{k,s}^{(4)} - 1\right) \leq Q^{rl} - Q_{k,s}^{avail,gen} \leq \mathbb{M} \cdot z_{k,s}^{(4)} - \epsilon \qquad \forall k \ \wedge \forall s \in \mathbb{S} \qquad \text{(A.18)}$$

$$\mathbb{M} \cdot \left(z_{k,s}^{(5)} - 1\right) \leq \hat{q}_k^r - Q_{k,s}^{avail,gen} \leq \mathbb{M} \cdot z_{k,s}^{(5)} - \epsilon \qquad \forall k \ \wedge \forall s \in \mathbb{S} \qquad \text{(A.19)}$$

$$-\mathbb{M} \cdot \left(1 + z_{k,s}^{(4)} - z_{k,s}^{(5)}\right) \leq q_{k,s}^{ract} - Q_{k,s}^{avail,gen} \leq \mathbb{M} \cdot \left(1 + z_{k,s}^{(4)} - z_{k,s}^{(5)}\right) \qquad \forall k \ \wedge \forall s \in \mathbb{S} \qquad \text{(A.20)}$$

$$-\mathbb{M} \cdot z_{k,s}^{(5)} \leq q_{k,s}^{ract} - \hat{q}_k^r \leq \mathbb{M} \cdot z_{k,s}^{(5)} \qquad \forall k \ \wedge \forall s \in \mathbb{S} \qquad \text{(A.21)}$$

### A.2.4. Receiver shutdown

Constraint ensures that a shutdown event occurs in the last step of the power generating mode and moves the receiver to off-mode in the next step. Constraints (A.22) prevents the receiver from power generating mode at time $k$ if a shutdown occurred in previous time $k-1$. Constraints (A.23) and (A.24) restrict the receiver's shutdown to power generation periods. The receiver is not able to follow the planned shutdown at time $k$ in scenario $s$ unless the receiver is at power generating mode at time $k$ in scenario $s$ (i.e., $\delta_{k,s}^r = 1$) and at the same time the dispatch instruction commands a planned shutdown (i.e., $\hat{y}_k^{rsd} = 1$), see Constraint (A.25).

$$\delta_{k-1,s}^{rsd} \geq \delta_{k-1,s}^r - \delta_{k,s}^r \qquad \forall k, \forall s \in \mathbb{S} \qquad \text{(A.22)}$$

$$\delta_{k-1,s}^{rsd} + \delta_{k,s}^r \leq 1 \qquad \forall k, \forall s \in \mathbb{S} \qquad \text{(A.23)}$$

$$\delta_{k,s}^{rsd} - \delta_{k,s}^r \leq 0 \qquad \forall k, \forall s \in \mathbb{S} \qquad \text{(A.24)}$$

$$-\mathbb{M} \cdot \left(2 - \hat{y}_k^{rsd} - \delta_{k,s}^r\right) \leq \delta_{k,s}^{rsd} - 1 \leq \mathbb{M} \cdot \left(2 - \hat{y}_k^{rsd} - \delta_{k,s}^r\right) \qquad \forall k, \forall s \in \mathbb{S} \qquad \text{(A.25)}$$

### A.2.5. Power block logical planned constraints

Constraints (A.26) to (A.29) impose relationships between binary planned variables associated with the power block. Constraint (A.26) counts the planned cold startup events. The shutdown event, however, is slightly different compared to the receiver such that the power block shutdown does not require thermal energy. This allows to exercise a shutdown event as soon as the power generation ends, see Constraint (A.27). Constraint (A.28) and (A.29) ensure the planned shutdown is performed only post power generating mode to transfer the power block to off-mode.

$$\hat{y}_k^{csup} \geq \hat{y}_k^c - \hat{y}_{k-1}^c \qquad \forall k \qquad \text{(A.26)}$$

$$\hat{y}_k^{csd} \geq \hat{y}_{k-1}^c - \hat{y}_k^c \qquad \forall k \qquad \text{(A.27)}$$

$$y_k^{csd} + y_k^c \leq 1 \qquad \forall k \qquad \text{(A.28)}$$

$$\hat{y}_k^{csd} \leq \hat{y}_{k-1}^c \qquad \forall k \qquad \text{(A.29)}$$

## A.2.6. Power block startup

There are two 0-1 auxiliary switches defined to drive power block's startup:

$$-\mathbb{M} \cdot \left(1 - z_{k,s}^{(6)}\right) \leq e_{k,s}^{csu} - E^c \leq \mathbb{M} \cdot z_{k,s}^{(6)} - \epsilon \qquad \forall k, \forall s \in \mathbb{S} \qquad \text{(A.30)}$$

$$\mathbb{M} \cdot \left(z_{k,s}^{(7)} - 1\right) \leq \Delta t \cdot Q^c - \varphi_{k,s} \leq \mathbb{M} \cdot z_{k,s}^{(7)} - \epsilon \qquad \forall k \ \wedge \forall s \in \mathbb{S} \qquad \text{(A.31)}$$

Constraints (A.32) to (A.36) are linear constraints expressing the conditions in Eq. (27):

$$\delta_{k,s}^{csu} \geq \hat{y}_k^c - \left(\delta_{k-1,s}^c + z_{k-1,s}^{(6)} + z_{k,s}^{(7)}\right) \qquad \forall k \wedge \forall s \in \mathbb{S} \qquad \text{(A.32)}$$

$$\delta_{k,s}^{csu} \leq \hat{y}_k^c \qquad \forall k \wedge \forall s \in \mathbb{S} \qquad \text{(A.33)}$$

$$\delta_{k,s}^{csu} + z_{k-1,s}^{(6)} \leq 1 \qquad \forall k \wedge \forall s \in \mathbb{S} \qquad \text{(A.34)}$$

$$\delta_{k,s}^{csu} + z_{k,s}^{(7)} \leq 1 \qquad \forall k \wedge \forall s \in \mathbb{S} \qquad \text{(A.35)}$$

$$\delta_{k,s}^{csu} + \delta_{k-1,s}^c \leq 1 \qquad \forall k \wedge \forall s \in \mathbb{S} \qquad \text{(A.36)}$$

## A.2.7. Power block power generation

Constraint (A.37) is a linear correlation for part-load power block operation, where $\dot{w}_{k,s}$ is the power block gross electricity generation, $\eta^p$ is the slope of linear approximation of the performance curve suggested by Wagner et al [46]; $Q^u$ is the maximum allowable input thermal power to the power block, and $W^u$ is the maximum electricity generation via the power block. Constraints (A.38) and (A.39) express the ramping up and down in the power block's operation at time $k$ in scenario $s$, respectively. Constraints (A.40) and (A.41) quantify the net electricity generation dispatched to the grid at time $k$ in scenario $s$ (i.e., $\dot{w}_{k.s}^{s}$) and the power purchased from the grid (i.e., $\dot{w}_{k,s}^{p}$) to satisfy the plant's parasitic losses, respectively.

$$\dot{w}_{k,s} = \eta^{p} \cdot q_{k,s}^{cact} + \delta_{k,s}^{c} \cdot (W^{u} - \eta^{p} \cdot Q^{u}) \qquad \forall k \ \wedge \forall s \in \mathbb{S} \qquad \text{(A.37)}$$

$$\dot{w}_{k,s}^{\delta} \geq \dot{w}_{k,s} - \dot{w}_{k-1,s} \qquad \forall k \ \wedge \forall s \in \mathbb{S} \qquad \text{(A.38)}$$

$$\dot{w}_{k,s}^{\delta} \geq \dot{w}_{k-1,s} - \dot{w}_{k,s} \qquad \forall k \ \wedge \forall s \in \mathbb{S} \qquad \text{(A.39)}$$

$$\dot{w}_{k.s}^{s} = \dot{w}_{k,s} \cdot (1 - \eta_{k}^{c}) \qquad \forall k \ \wedge \forall s \in \mathbb{S} \qquad \text{(A.40)}$$

$$\begin{aligned} \dot{w}_{k,s}^{p} = {} & L^{r} \cdot \left(q_{k,s}^{ract} + Q^{ru} \cdot \delta_{k,s}^{rsu}\right) + L^{c} \cdot q_{k,s}^{cact} \\ & + W^{h} \cdot \delta_{k,s}^{r} + \frac{E^{hs}}{\Delta t} \cdot \left(\delta_{k,s}^{rsd} + \delta_{k,s}^{rsu}\right) \end{aligned} \qquad \forall k \ \wedge \forall s \in \mathbb{S} \qquad \text{(A.41)}$$

### A.2.8. Power block shutdown

The power block shutdown is because of a planned shutdown from the dispatch instruction, or a forced shutdown due to insufficient storage. Constraint (A.42) forces a planned shutdown at time $k$ and scenario $s$ provided that the power block was under power generating mode in the previous time $k-1$ in scenario $s$. With Constraint (A.43), the shutdown only occurs at time $k$ in scenario $s$ post when power block stops electricity generation. Constraint (A.44) ensures shutdown and power generation are two mutually exclusive events and never occur simultaneously.

$$-\mathbb{M} \cdot \left(2 - \hat{y}_{k}^{csd} - \delta_{k-1,s}^{c}\right) \leq \delta_{k,s}^{rsd} - 1 \leq \mathbb{M} \cdot \left(2 - \hat{y}_{k}^{rsd} - \delta_{k-1,s}^{r}\right) \qquad \forall k \ \wedge \forall s \in \mathbb{S} \qquad \text{(A.42)}$$

$$\delta_{k,s}^{csd} \geq \delta_{k-1,s}^{c} - \delta_{k,s}^{c} \qquad \forall k \ \wedge \forall s \in \mathbb{S} \qquad \text{(A.43)}$$

$$\delta_{k,s}^{csd} + \delta_{k,s}^{c} \leq 1 \qquad \forall k \ \wedge \forall s \in \mathbb{S} \qquad \text{(A.44)}$$

# Appendix B. Supplementary Materials

## B.2 Dispatch planning using Heuristic-2

In this section the results of dispatching under Heuristic-2 for the illustrative example. Figure 12 and Figure 13 show the optimized DP and the actual variables in the receiver's and the power block's operation, respectively.

**Figure 12: The receiver's plan and control variables under Heurisitc 2**

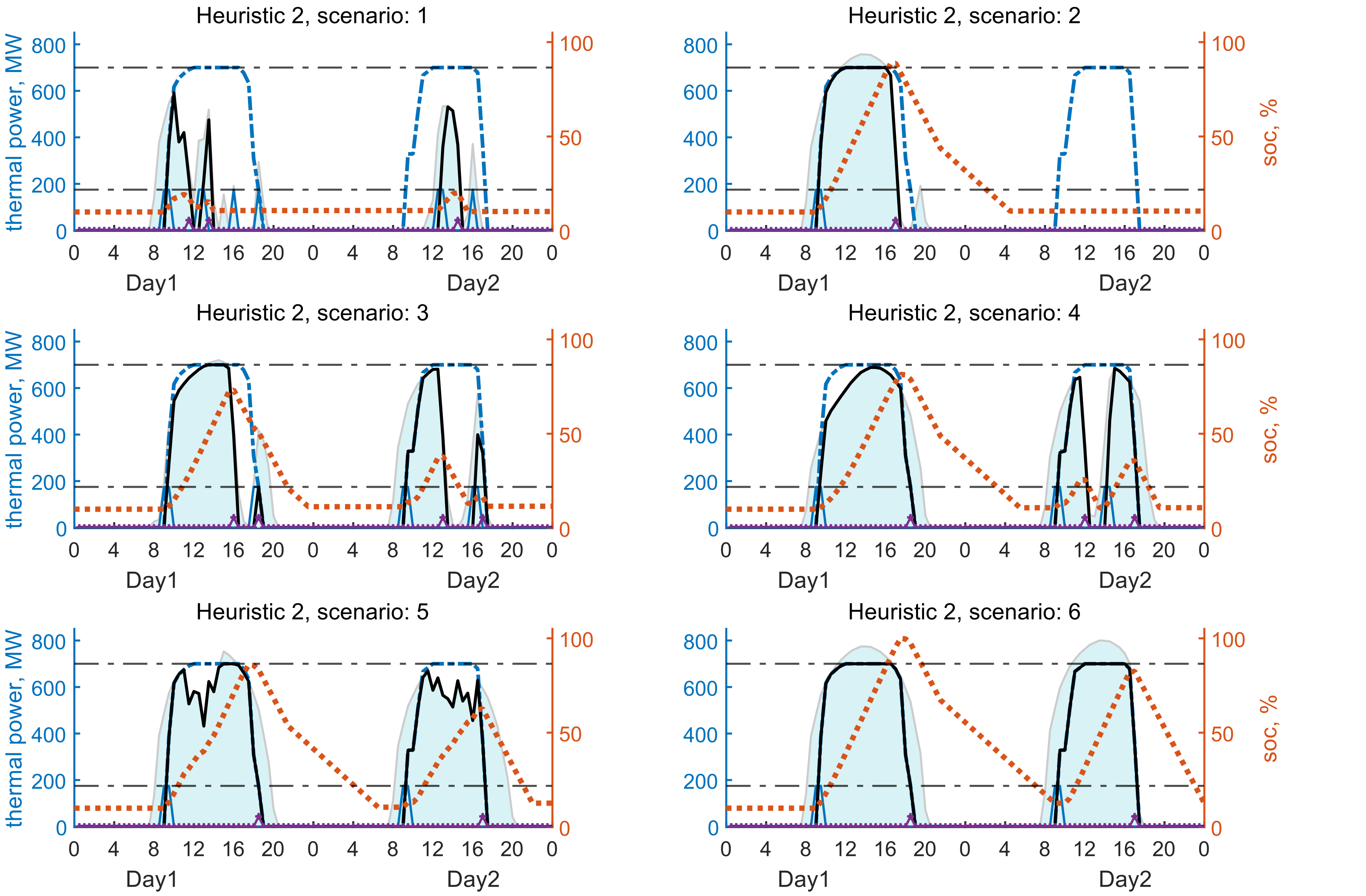

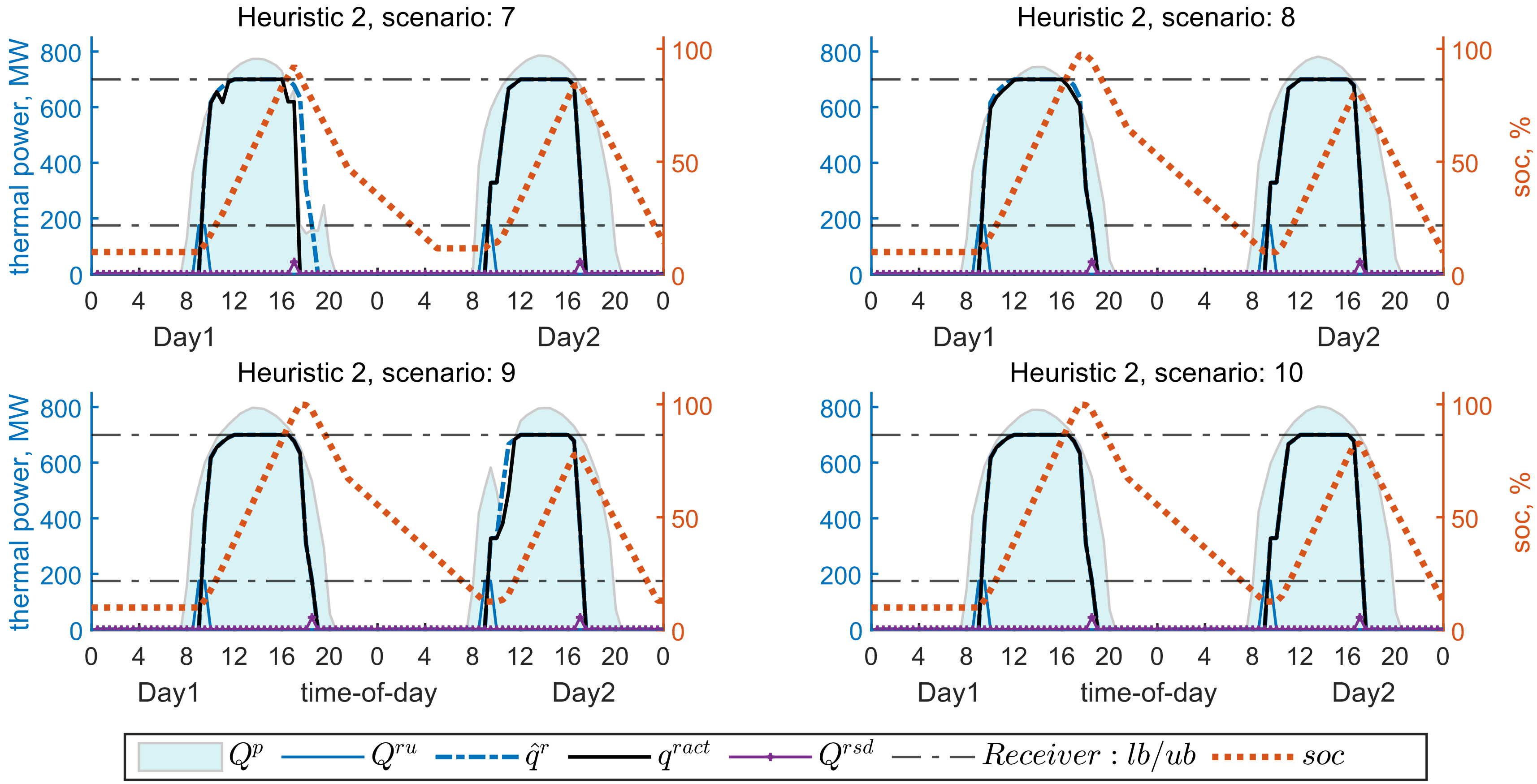
Heuristic 2, scenario: 7
Heuristic 2, scenario: 8
Heuristic 2, scenario: 9
Heuristic 2, scenario: 10
thermal power, MW
soc, %
0 4 8 12 16 20 0 4 8 12 16 20 0
Day1
time-of-day
Day2
$Q^p$
$Q^{ru}$
$\hat{q}^r$
$q^{ract}$
$Q^{rsd}$
$Receiver : lb/ub$
$soc$

**Figure 13: The power block (PB)'s planned and control variables under Heuristic-2.**

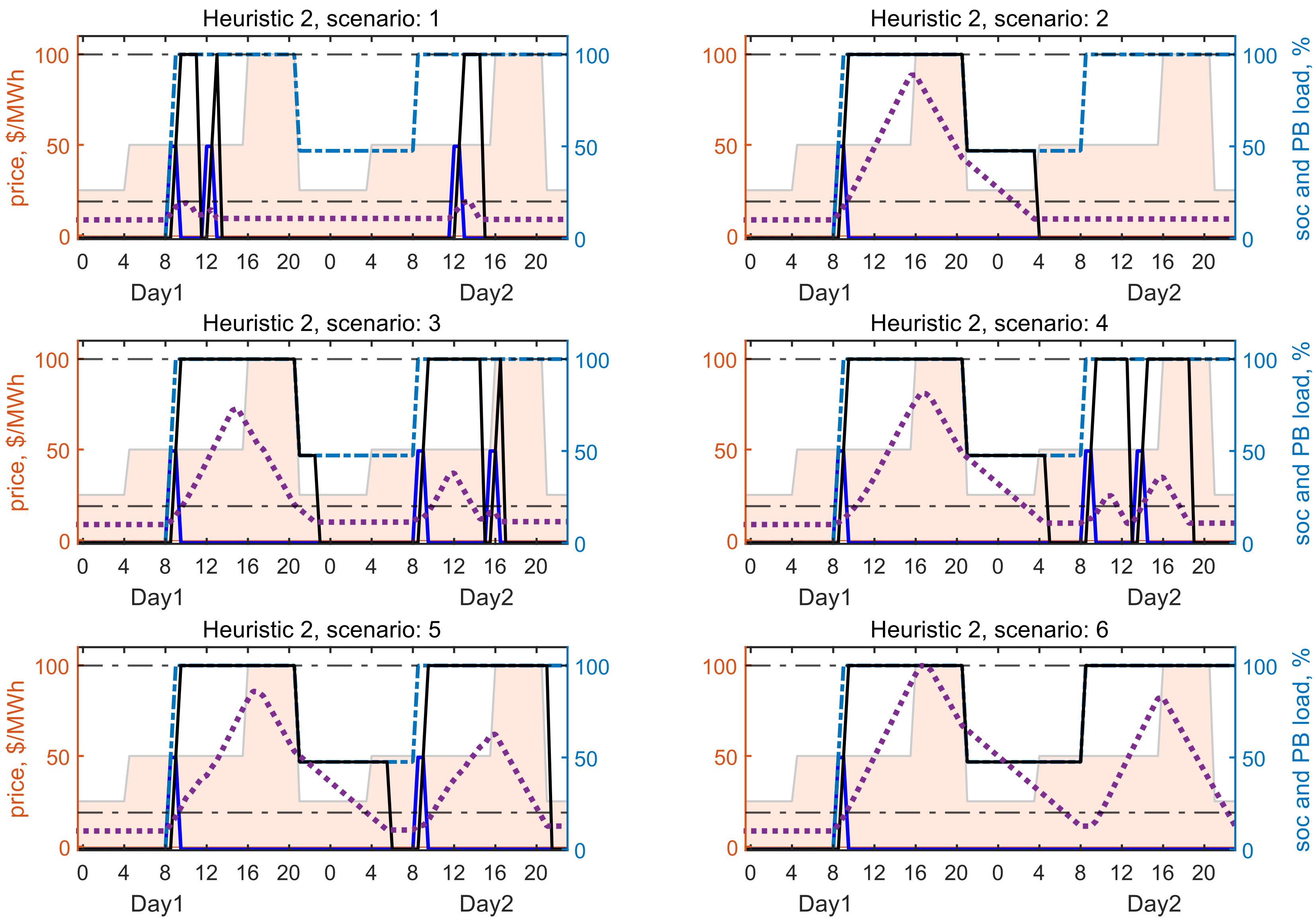

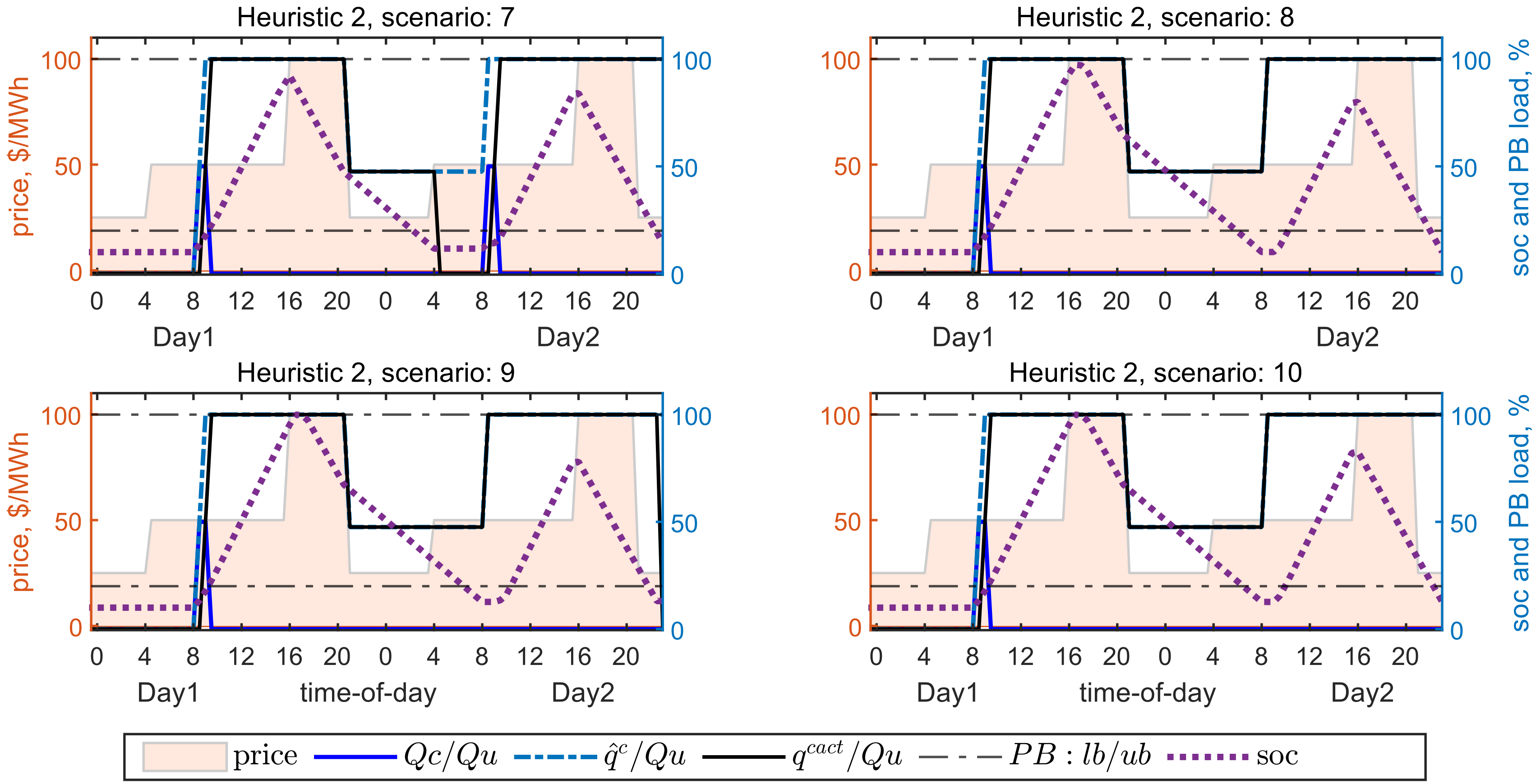
Heuristic 2, scenario: 7
Heuristic 2, scenario: 8
Heuristic 2, scenario: 9
Heuristic 2, scenario: 10
price, $/MWh
soc and PB load, %
0 4 8 12 16 20 0 4 8 12 16 20
Day1
time-of-day
Day2
price
Qc/Qu
q̂^c/Qu
q^cact/Qu
PB : lb/ub
soc